\documentclass[a4paper,fleqn,usenatbib]{mnras_tex_edited}

\usepackage{amssymb,amsmath,verbatim,mathtools,needspace,enumitem,etoolbox,graphicx,physics,microtype,natbib,url,hyperref,mathrsfs,bm,ulem}

\newcommand{\ssim}{\mathchar"5218\relax\,}

\renewcommand{\emph}[1]{\textit{#1}}

\newcommand{\approptoinn}[2]{\mathrel{\vcenter{
  \offinterlineskip\halign{\hfil$##$\cr
    #1\propto\cr\noalign{\kern2pt}#1\sim\cr\noalign{\kern-2pt}}}}}

\def\araa{\rm{ARA\&A}}
\def\apj{\rm{ApJ}}
\def\apjl{\rm{ApJ}}
\def\apjs{\rm{ApJS}}
\def\aap{\rm{A\&A}}
\def\aapr{\rm{A\&A~Rev.}}

\def\mnras{\rm{MNRAS}}
\def\prd{\rm{PRD}}
\def\prl{\rm{PRL}}

\def\ssr{\rm{Space~Sci.~Rev.}}
\def\nat{\rm{Nature}}

\def\cqg{\rm{CQG}}

\title[Bardeen-Petterson effect in SMBH binaries]{The Bardeen-Petterson effect in accreting supermassive black-hole binaries: a systematic approach}

\newcommand{\bham}{School of Physics and Astronomy \&	 Institute for Gravitational Wave Astronomy, University of Birmingham,\vspace{-0.05cm}\\$\;$Birmingham, B15 2TT, UK}
\newcommand{\leiden}{Leiden Observatory, Leiden University, P.O.~Box 9513, NL-2300~RA Leiden, the Netherlands}
\newcommand{\aei}{Max Planck Institute for Gravitational Physics (Albert Einstein Institute), Am M\"uhlenberg 1, Potsdam 14476, Germany}

\author[Gerosa et al.]{Davide Gerosa$^1$\thanks{\href{mailto:d.gerosa@bham.ac.uk}{d.gerosa@bham.ac.uk}}, Giovanni Rosotti$^2$, Riccardo Barbieri$^3$
\bigskip \\
$^1$ \bham\\
$^2$ \leiden\\
$^3$ \aei
}
\pubyear{2020}

\begin{document}

\label{firstpage}
\pagerange{\pageref{firstpage}--\pageref{lastpage}}
\maketitle

\begin{abstract}
Disc-driven migration is a key evolutionary stage of supermassive black-hole binaries hosted in gas-rich galaxies. Besides promoting the inspiral, viscous interactions tend to align the spins of the black holes with the orbital angular momentum of the disc. We present a critical and systematic investigation of this problem, also known as the Bardeen-Petterson effect. 
We design a new iterative scheme to solve the non-linear dynamics of warped accretion discs under the influence of both relativistic frame dragging and binary companion. {We characterize the impact of the disc ``critical obliquity'', which marks regions of the parameter space where stationary solutions do not exist.} We find that black-hole spins reach either complete alignment or a critical configuration. {Reaching the critical obliquity might imply that the disc breaks as observed in hydrodynamical simulations.} %
Our findings are important to predict the spin configurations with which supermassive black-hole binaries enter their gravitational-wave driven regime and become detectable by~LISA.
\end{abstract}


\section{Introduction}

 Cosmological probes and observations of interacting galaxies all point to a scenario where structures grow hierarchically. Supermassive black holes (BHs) are believed to form binaries and merge with each other following the mergers of their host galaxies \citep{2014ARA&A..52..589H}. Several observational candidates of binary supermassive BHs have been reported to date with signatures spanning from blazars with quasi-periodic outbursts \citep{1996ApJ...460..207L}, dual AGNs \citep{2003ApJ...582L..15K,2015ApJ...806..219C}, compact radio cores \citep{2006ApJ...646...49R,2017NatAs...1..727K}, and quasars with either optical variability \citep{2015MNRAS.453.1562G,2016MNRAS.463.2145C} or spectroscopically distinct features \citep{2012ApJS..201...23E}. During their late inspiral and merger phase, supermassive BHs emit copious gravitational waves at frequencies targeted by the LISA space mission \citep{2017arXiv170200786A} and Pulsar Timing Arrays \citep{2019A&ARv..27....5B}. %

The pairing process of supermassive BHs is one of the most outstanding problems in high-energy astrophysics \citep{1980Natur.287..307B} ---for a review see  \cite{2014SSRv..183..189C}. Following a galaxy merger, the two  BHs are first brought together by dynamical friction and star scattering, which decrease the binary separation down to about 1 pc. Gravitational-wave emission, however, can successfully drive the inspiral only from a much smaller separation of $\ssim 10^{-3}$ pc. A variety of processes have been invoked to bridge these two regimes in what has been dubbed as the \emph{final parsec problem} \citep{2001ApJ...563...34M}. These include triaxial galactic
potentials \citep{2004ApJ...606..774P}, dynamical interactions in supermassive BH triples \citep{2019MNRAS.486.4044B}, and,
crucially to the scope of this paper, gas accretion (e.g. \citealt{2002ApJ...567L...9A,2005ApJ...630..152E,2009ApJ...700.1952H,2009MNRAS.398.1392L,2012A&A...545A.127R,2013CQGra..30x4008M,2017MNRAS.469.4258T}). Most likely, a combination
of all these is at play in the Universe, with different processes being more or less relevant for specific type of hosts and BHs.

For gas-rich environments, disc accretion  provides a natural way to facilitate the merger. This process is analogous to planetary migration \citep{1986ApJ...309..846L}, which is commonly invoked to explain the presence of giant planets close to their central stars \citep{1996Natur.380..606L}: the gravitational interaction with the disc transfers angular momentum, in general in a direction going from the binary to the disc, and the orbital separation consequently shrinks.
Gas accretion leaves a deep imprint on the assembly history of BH binaries that could potentially be reconstructed by future gravitational-wave observations (e.g. \citealt{2008ApJ...684..822B,2011PhRvD..83d4036S,2017MNRAS.464.3131K,2017PhRvL.118r1102T}).

When embedded in circumbinary discs, the BHs carve a cavity (or a gap, depending on the BH masses) around the binary \citep{1980ApJ...241..425G}. Mass streams from the circumbinary disc penetrate the cavity, forming smaller individual discs (also called \emph{minidiscs}, or \emph{circum-BH discs}) around the two BHs  %
\citep{1996ApJ...467L..77A,2014ApJ...783..134F,2017ApJ...838...42B,2018ApJ...853L..17B}. 
In general, the BH spins and their discs will not share the same orientation. This is especially true in a scenario where BHs were brought together by many, randomly oriented stellar encounters during the previous phase of their evolution.

In such a setup, gas accretion will have a deep impact on the spin orientations. The process is known as the \emph{Bardeen-Petterson effect} and is due to a combination of general-relativistic frame dragging and viscous interactions \citep{1975ApJ...195L..65B,1978Natur.275..516R,1985MNRAS.213..435K}.   The inner disc (up to the so-called ``warp radius'') aligns to the BH equatorial plane on the short viscous timescale. The outer disc, which contains most of the angular momentum, maintains its initially tilted orientation and reacts by pulling the BH towards complete alignment on a longer timescale of~$\ssim 10^6$ yr.

As spin alignment takes place, the disc presents a non-planar, warped structure \citep{1996MNRAS.282..291S,2007MNRAS.381.1617M,2009MNRAS.399.2249P}. At the warp radius, the mass surface density might drop by several orders of magnitude \citep{2014MNRAS.441.1408T}, potentially reducing the effectiveness of the Bardeen-Petterson effect. In this regime, warp propagation is non-linear and the fluid viscosities depend on the details of the disc profile \citep{1999MNRAS.304..557O,2013MNRAS.429L..30L,2013MNRAS.433.2403O}. The disc of each BH is subject to the additional perturbation of the binary companion \citep{2009MNRAS.400..383M,2010MNRAS.402..682D}, which   pushes the warp radius inwards and speeds up the alignment \citep{2013ApJ...774...43M}.  Furthermore, \cite{2014MNRAS.441.1408T} reported the presence of a ``critical obliquity'' where viable disc profiles cease to exist if the inclination of the disc is too high. %

In this paper, we put together all these ingredients for the first time, presenting a new, systematic approach to the Bardeen-Petterson effect in supermassive BH binaries. Depletion of the surface density, non-linear warp propagation, perturbation of the BH companion, and critical obliquity all play a crucial role in determining the mutual orientations of BHs and their discs. Most previous works only focused on determining the disc shape and not on the role of these effects on the spin-alignment process. This study is an important step to go beyond timescale comparisons \citep{2007ApJ...661L.147B,2013MNRAS.429L..30L,2013ApJ...774...43M,2015MNRAS.451.3941G} and predict the residual spin orientations supermassive BH binaries are left with following their disc-driven phase. A future publication will explore the relevance of our findings to gravitational-wave observations.

This paper is organized as follows. In Sec.~\ref{warpeddiscs}, we present the equations of warped accretion discs subject to the perturbation of both relativistic frame dragging and the BH companion. In Sec.~\ref{sectowards}, we design and test a new iterative scheme to capture the effect of non-linear warp propagation. In Sec.~\ref{BPbin},  we present a detailed study of the Bardeen-Petterson effect in binaries, highlighting the importance of the shape of the disc and the critical obliquity. In Sec.~\ref{secjoint}, we present a preliminary investigation of the coupled evolution of BH spin alignment and gas-driven migration. Finally, in Sec.~\ref{secconcl}, we discuss relevance and limitations of our findings.

\section{Warped accretion discs}
\label{warpeddiscs}

We first write down the equations  governing the dynamics of warped accretion discs and reduce them to dimensionless variables.

\subsection{Evolutionary equations}

Let us consider a disc surrounding a BH of mass $M$ and spin $\mathbf{J}=  G \chi M^2 \mathbf{\hat J}/c$, where $\chi\in[0,1]$ is the dimensionless Kerr parameter. The disc is modeled as a superposition of rings at a distance $R$ from the BH. The surface mass density of the disc is denoted by $\Sigma$ and the angular momentum of each ring is denoted by $\mathbf{L}$. We assume Keplerian discs, i.e. $L =\Sigma \sqrt{GMR}  $. The BH is orbiting a companion of mass $M_\star$; the separation and angular momentum of the binary are denoted by $R_\star$ and  $\mathbf{L}_\star$, respectively. Is it also useful to define the warp amplitude $\psi = R |\partial\mathbf{\hat L}/\partial R|$.

The dynamics of the disc is set by mass and momentum conservation \citep{1983MNRAS.202.1181P,1985MNRAS.213..435K,1992MNRAS.258..811P,1999MNRAS.304..557O,2001MNRAS.320..485O,2007MNRAS.381.1617M,2009MNRAS.400..383M,2013MNRAS.433.2403O}:
\begin{align}
&\frac{\partial\Sigma}{\partial t} = \frac{3}{R} \frac{\partial}{\partial R}\left[ R^{1/2}\frac{\partial}{\partial R}\left( \nu_1 \Sigma R^{1/2} \right)\right]
+ \frac{1}{R}\frac{\partial}{\partial R}\left[  \nu_2 \Sigma R^2 \left| \frac{\partial \mathbf{\hat L}}{\partial R}\right|^2\right]\,,
\label{pringlemass}
\\
&{\frac{\partial \mathbf{L}}{\partial t}= 
\frac{3}{R} \frac{\partial}{\partial R} \left[ \frac{R^{1/2}}{\Sigma}  \frac{\partial}{\partial R} \left( \nu_1 \Sigma R^{1/2} \right) \mathbf{ L} \right]
+\frac{1}{R} \frac{\partial}{\partial R}\Bigg[ \Bigg( \nu_2 R^2 \left| \frac{\partial \mathbf{\hat L}}{\partial R}\right|^2}
\notag \\
&
{ - \frac{3}{2}\nu_1 \Bigg)  \mathbf{ L}\Bigg] + \frac{1}{R} \frac{\partial}{\partial R} \left( \frac{1}{2} \nu_2 R  L \frac{\partial \mathbf{\hat L}}{\partial R} \right) 
+\frac{\partial}{\partial R} \left( \nu_3 R  \mathbf{L} \cross \frac{\partial \mathbf{\hat L}}{\partial R}\right) }
\notag \\
&{ + \frac{2G}{c^2}  \frac{\mathbf{J} \times \mathbf{ L}}{R^3} 
 + \frac{3GM_\star \Sigma R^2}{4 R_\star^3} 
 \left(\mathbf{\hat L} \cdot \mathbf{\hat L}_\star\right) \left(\mathbf{\hat L} \times \mathbf{\hat L}_\star\right)\,.}
 \label{pringlemom}
\end{align}
The viscosity $\nu_1$ models the response of the disc to azimuthal stresses associated with disc accretion. The viscosity $\nu_2$ models the vertical resistance of the disc to be warped.  The precession contribution proportional to $\nu_3$ does not impact the disc dynamics  \citep{2010MNRAS.405.1212L,2014MNRAS.441.1408T} and is here neglected.  The torque proportional to $(\mathbf{J} \times \mathbf{L})$ models Lense-Thrirring precession and is responsible for aligning the inner disc with with the BH spin. The term proportional to $(\mathbf{\hat L} \cdot \mathbf{\hat L}_\star) (\mathbf{\hat L} \times \mathbf{\hat L}_\star)$ models the external torque imparted by the companion and is responsible for aligning the outer disc with the binary's orbital plane. 
We will solve this set of equation imposing that the inner and outer disc are aligned with $\mathbf{J}$ and $\mathbf{L}_\star$, respectively.

Let us focus on steady-state solutions, i.e. $\partial\Sigma/\partial t=0$ and $\partial\mathbf{L}/\partial t=0$. Equation~(\ref{pringlemass}) can be integrated to obtain  %
\begin{align}
3 R^{1/2}\frac{\partial}{\partial R}\left( \nu_1 \Sigma R^{1/2} \right)
+ \nu_2 \Sigma R^2 \left| \frac{\partial \mathbf{\hat L}}{\partial R}\right|^2 = \frac{\dot M}{2 \pi}\,
\label{masscurrent}
\end{align}
where the constant $\dot M$ is positive for mass flowing onto the BH. The accretion rate  can be conveniently parametrized as 
\begin{align}
\dot M = f \frac{M}{t_{\rm Edd}}
\label{eddfraction}
\end{align}
where $t_{\rm Edd} \simeq 4.5 \times 10^8$ yr \citep{1964ApJ...140..796S}. The Eddington limit corresponds to $f=1/\epsilon$, where $\epsilon=\epsilon(\chi)\ssim 0.1$ is the accretion efficiency \citep{1973blho.conf..215B}. 
Equation~(\ref{masscurrent}) reduces to the familiar limit $\dot M= 3 \pi \nu_1 \Sigma $ for planar discs at large radii (e.g.~\citealt{2002apa..book.....F,2008NewAR..52...21L}). Setting $\nu_3=0$, Eq.~(\ref{pringlemom}) yields
{
\begin{align}
&\frac{1}{2} \frac{\partial }{\partial R}\left(\dot M \pi R^{1/2} \mathbf{\hat L} - 3 \nu_1 R^{1/2} \Sigma \mathbf{\hat L} + \nu_2 R^{3/2}  \Sigma \frac{\partial \mathbf{\hat L}}{\partial R} \right) 
\notag \\&+  \frac{2G^2 M^2 \chi \Sigma}{c^3 R^{3/2} }  \left(\mathbf{\hat J} \times \mathbf{\hat L}\right)
 \!+\! \frac{3  \Sigma  R^3 GM_\star  }{4 R_\star^3 \sqrt{GM}}\! \left(\mathbf{\hat L} \cdot \mathbf{\hat L}_\star\right) \!\left(\mathbf{\hat L} \times \mathbf{\hat L}_\star\right) \!=0.
 \label{momcurrent}
\end{align}}

We use the \cite{1973A&A....24..337S} prescription and parametrize the viscosities in terms of dimensionless coefficients $\alpha_1$ and $\alpha_2$. In particular, we assume \citep{2007MNRAS.381.1617M,2009MNRAS.400..383M}
\begin{align}
\nu_1 = \nu_0 \left(\frac{R}{R_0}\right)^\beta \alpha_1(\alpha,\psi) \label{nu1}
\\
\nu_2= \nu_0  \left(\frac{R}{R_0}\right)^\beta \alpha_2(\alpha,\psi)
\label{nu2}
\end{align}
where $\nu_0$, $R_0$, and $\beta$ are constant. 
 In general,  $\alpha_1$ and $\alpha_2$ are functions of both the kinematic viscosity parameter $\alpha$ and the warp amplitude $\psi$ \citep{1999MNRAS.304..557O,2013MNRAS.433.2403O}. In the small-warp limit one has\footnote{\cite{2013MNRAS.433.2403O} make use of the equivalent notation $Q_1=-3 \alpha_1/2$ and {$Q_2=\alpha_2/2$} (which is not identical to the notation used by \citealt{1999MNRAS.304..557O}; see also \citealt{2018MNRAS.476.1519D}).}
\begin{align}
\label{alpha1og}
\lim_{\psi \to 0} \alpha_1(\alpha,\psi) &=  \alpha +\mathcal{O}(\psi^2)
\\ 
\lim_{\psi \to 0} \alpha_2(\alpha,\psi)&= \frac{2(1+7\alpha^2)}{\alpha(4+\alpha^2)} +\mathcal{O}(\psi^2)\,.
\label{alpha2og}
\end{align}
such that Eq.~(\ref{nu1}) reduces to the usual expression $\nu_1\propto \alpha R^\beta$ (e.g.~\citealt{2002apa..book.....F,2008NewAR..52...21L}). The viscosity is related to the temperature by
\begin{align}
\nu_0 = \left(\frac{H_0}{R_0}\right)^2 \sqrt{GM R_0}
\label{nu0HR}
\end{align}
where $H_0$ is the vertical height of the disc at $R_0$.

In this paper, we use the isothermal theory of \citet{2013MNRAS.433.2403O} to set the values of the viscosity coefficients. It is worth stressing that in this context isothermal (as opposed to adiabatic) refers to how the disc responds to a perturbation, i.e. it should be intended as \emph{locally} isothermal. Using the isothermal theory for the viscosity coefficient does not mean that we are restricted to study globally isothermal discs, which are described by $\beta=3/2$. In what follows, we will also explore different values of $\beta$, which imply that the temperature is a function of radius. %

Let us further define  
\begin{align}
\tilde\alpha_i(\alpha,\psi) = \frac{\alpha_i(\alpha,\psi)}{\alpha_i(\alpha,\psi\!=\!0)}  \qquad {\rm for}\quad i\!=\!1,2
\end{align}
 and 
\begin{align}\zeta = \frac{\alpha_2(\alpha,\psi\!=\!0)}{\alpha_1(\alpha,\psi\!=\!0)} =\frac{2(1+7\alpha^2)}{\alpha^2(4+\alpha^2)} \,.
\end{align}
For $\alpha \to 0$ and $\psi\to0$ one obtains the leading-order  expression {$\zeta \simeq 1/2\alpha^2$} \citep{1983MNRAS.202.1181P}.

\begin{figure}\centering
\includegraphics[width=0.9\columnwidth]{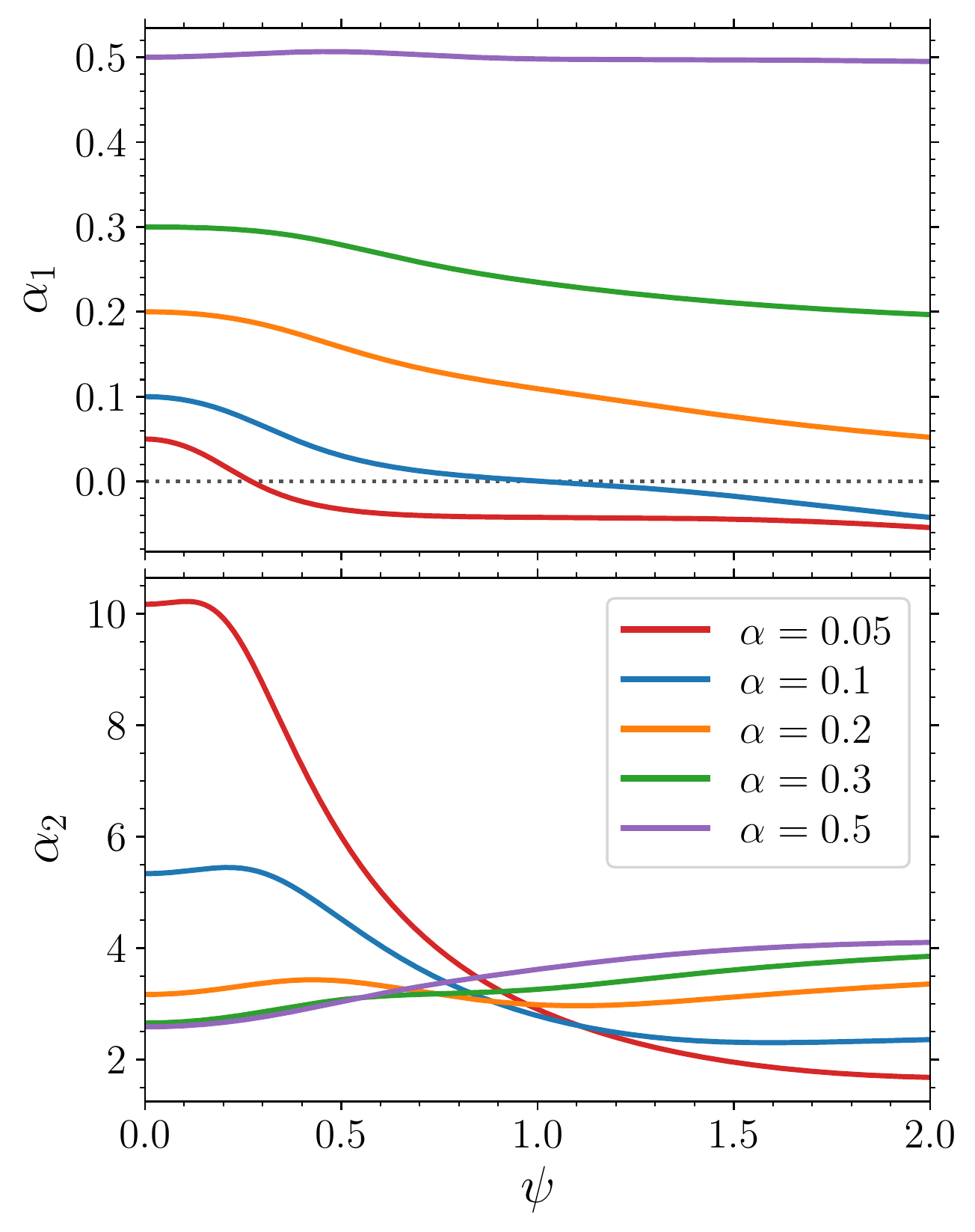}
\caption{Horizontal ($\alpha_1$, top panel) and vertical ($\alpha_2$, bottom panel) viscosity coefficients as a function of  warp amplitude $\psi = R |\partial{\mathbf{\hat L}}/\partial R|$ and  Shakura-Sunyaev parameter $\alpha$. We assume the disc is locally isothermal {and compute the viscosities following \protect\cite{2013MNRAS.433.2403O}}.}
\label{alphavspsi}
\end{figure}

Figure~\ref{alphavspsi} shows the behaviour of $\alpha_1$ and $\alpha_2$ as a function of $\psi$ and  $\alpha$ {computed following \cite{2013MNRAS.433.2403O}}.   For $\alpha\lesssim 0.1$, the viscosity coefficients decrease with the warp amplitude. In particular, $\alpha_1$ becomes negative at moderate values $\psi\lesssim 1$ \citep{2018MNRAS.476.1519D}. As explored at length in the sections below, this viscosity regime might cause a sharp breaking of the disc \citep{2012MNRAS.421.1201N,2013MNRAS.434.1946N,2015MNRAS.448.1526N} which is not captured by our integrations.

\subsection{Dimensionless variables}

We now rewrite the disc equations using dimensionless variables.
We scale the radial coordinate $R$ with the constant $R_0$ appearing in Eqs.~(\ref{nu1}-\ref{nu2}) and the surface density $\Sigma$ with the accretion rate $\dot M$ and the viscosity $\nu_0$. More specifically, we define %
\begin{align}
r=\frac{R}{R_0}\,, \qquad \sigma= \frac{2 \pi}{\dot M} \alpha\nu_0 \Sigma\,;
\label{dimlessscaling}
\end{align}
where the numerical factors have been chosen for consistency with \cite{2014MNRAS.441.1408T}.  Equations~(\ref{masscurrent}-\ref{momcurrent}) can be rewritten as
\begin{align}
\frac{\partial \sigma}{\partial r} &=   - \left( \beta  + \frac{1}{2}\right) \frac{\sigma}{r} - \frac{\zeta \sigma \psi^2}{3r} \frac{\tilde\alpha_2(\alpha,\psi)}{ \tilde\alpha_1(\alpha,\psi)}
\notag \\ &+ \frac{r^{-\beta -1}}{3 \tilde\alpha_1(\alpha,\psi)} - \frac{\sigma}{\tilde\alpha_1(\alpha,\psi)} \frac{\partial \tilde\alpha_1(\alpha,\psi)}{\partial r} \,,
\label{tildealpha1}
\end{align}
\begin{align}
\frac{\partial^2 \mathbf{\hat L}}{\partial r^2} &=  \frac{\partial \mathbf{\hat L}}{\partial r} \bigg[-\frac{2 r^{-\beta-1} }{\zeta\, \tilde\alpha_2(\alpha,\psi) \sigma} + \frac{3}{\zeta r}\frac{ \tilde\alpha_1(\alpha,\psi)}{ \tilde\alpha_2(\alpha,\psi)} 
- \left(\beta+\frac{3}{2}\right)\frac{1}{r} 
\notag \\& 
- \frac{1}{\sigma}\frac{\partial \sigma}{\partial r} - \frac{1}{\tilde\alpha_2(\alpha,\psi)}\frac{\partial \tilde\alpha_2(\alpha,\psi)}{\partial r} \bigg] 
 -  \frac{\psi^2}{r^2}  \mathbf{\hat L}
 \notag \\&
-  \left(\frac{R_{\rm LT}}{R_0}\right) \frac{r^{-\beta-3}}{\tilde\alpha_2(\alpha,\psi)} \left(\mathbf{\hat J} \times \mathbf{\hat L}\right)
\notag \\ &-\left(\frac{R_{\rm tid}}{R_{0}}\right)^{-7/2}  \frac{ r^{-\beta+3/2}}{\tilde\alpha_2(\alpha,\psi)}  \left(\mathbf{\hat L} \cdot \mathbf{\hat L}_\star\right) \left(\mathbf{\hat L} \times \mathbf{\hat L}_\star\right).
\label{tildealpha2}
\end{align}
where 
\begin{align}
R_{\rm LT} &= \frac{4 G^2 M^2 \chi}{c^3 \alpha \nu_0 \zeta}\,,
\label{rltdef}
\\
{R_{\rm tid}} & {=\left( \frac{2}{3} \frac{\sqrt{GM}}{GM_\star} R_\star^3 \alpha\nu_0 \zeta \right)^{2/7}\,.}
\label{rtiddef}
\end{align}

{The quantities $R_{\rm LT}$ and $R_{\rm tid}$ mark the typical location in the disc where Lense-Thirring and tidal external torques, respectively, mostly affect the warp profile}~\citep{2009MNRAS.400..383M}.
It is convenient to measure the viscosities in Eqs.~(\ref{nu1}-\ref{nu2}) from either of these two lenghtscales where the warp is expected to be large, such that solutions for different values of $\beta$ can be compared meaningfully. In particular, we set 
\begin{align}
R_0=R_{\rm LT}
\end{align}
such that the evolutionary equations depend only on the dimensionless parameter
\begin{align}
\kappa= \left(\frac{R_{\rm tid}}{R_{\rm LT}}\right)^{-7/2}\,.
\label{kappadef}
\end{align}

It is useful to combine Eqs.~(\ref{rtiddef}) and (\ref{nu0HR}) into
\begin{equation}
\nu_0 = \frac{GM}{c} \left(\frac{H_0}{R_0}\right)^{4/3} \left(\frac{4 \chi}{\alpha\zeta}\right)^{1/3}\,.\label{fullnu0}
\end{equation}
to obtain
\begin{align}
&R_{\rm LT}\simeq 1.6 \times 10^{-3} 
\left(\frac{M}{10^7 M_\odot} \right)
\left(\frac{\chi}{0.5} \right)^{2/3} 
\left(\frac{H_0/R_0}{0.002} \right)^{-4/3}
\notag \\&\quad\;\;\times
\left(\frac{\alpha}{0.2} \right)^{-2}
\left[\frac{\zeta}{1/(2\!\times\! 0.2^2)} \right]^{-2} {\rm pc}\,,
\\
&{\kappa\simeq 0.66
\left(\frac{M}{10^7 M_\odot} \right)^2
\left(\frac{\chi}{0.5} \right)^{2} 
\left(\frac{M_\star}{10^7 M_\odot} \right)
\left(\frac{R_\star}{0.1 {\rm pc}} \right)^{-3}}
\notag \\&\;\;{\times
\left(\frac{H_0/R_0}{0.002} \right)^{-6}
\left(\frac{\alpha}{0.2} \right)^{-3}
\left[\frac{\zeta}{1/(2\!\times\! 0.2^2)} \right]^{-3}\,,}
\label{kappavalue}
\end{align}
where we used {$\zeta \simeq 1/2\alpha^2$} to set a fiducial value for $\zeta$. 

If $\kappa=0$, the effect of the companion is negligible and the system reduces to that of a single BH and its surrounding accretion disc. In this case, the solution is self-similar \citep{2007MNRAS.381.1617M}: a more massive or more rapidly spinning BH would be surrounded by a scaled-up disc with a larger warp radius but identical shape. This is not the case for $\kappa\neq 0$, where the relative importance of torques imparted by relativistic frame dragging and the binary companion plays a crucial role. As an example, note how $\kappa$ depends separately on masses of the two BHs, and not only on their ratio.

\subsection{Numerical setup}
\label{numsetup}

We solve Eq.~(\ref{tildealpha1}-\ref{tildealpha2}) as a first-order boundary value problem (BVP) for $\sigma$,  $\mathbf{\hat L}$, and $\partial \mathbf{\hat L}/ \partial r$. {Numerical implementations treat the cartesian components of  $\mathbf{\hat L}
$ and $\partial \mathbf{\hat L}/ \partial r$ as independent variables; the constraint $|\mathbf{\hat L}|=1$ must be imposed with suitable boundary conditions.}  

We use a 4-th order collocation algorithm as implemented in \texttt{scipy.integrate.solve\_bvp} \citep{2019arXiv190710121V} with a tolerance of $10^{-3}$ and a radial grid ranging from  $r_{\rm min}=10^{-1}$ to $r_{\rm max}=10^4$. {We initialize our numerical grid with 500 nodes equispaced in log between $r_{\rm min}$ and $r_{\rm max}$. The algorithm then add gridpoints if and where it is deemed necessary to reach the targeted tolerance. Converged solutions typically present $\lesssim 1000$ gridpoints}

We assume a reference frame where the BH spin lies along the $z$-axis and the binary orbital angular momentum lies in the $xz$-plane, i.e.
\begin{align}
\mathbf{\hat J}=(0,0,1)\,,
\qquad
\mathbf{\hat L}_\star=(\sin\theta,0,\cos\theta)\,.
\label{coordass}
\end{align}
The angle $\theta$ parametrizes the misalignment between the BH spin and the outer disc.

Our BVP requires seven boundary conditions:
\begin{itemize}[leftmargin=0.6cm]
\item
We assume that the binary angular momentum tracks the direction of the mass inflow at large separations, i.e. 
\begin{align}
\mathbf{\hat L}(r_{\rm max}) =\mathbf{\hat{L}}_\star\,,
\label{BCbin}
\end{align}
which corresponds to three constraints.
\item
At the outer edge of the grid we impose
\begin{align}
{\mathbf{\hat L}(r_{\rm max})\cdot \frac{\partial \mathbf{\hat L}}{\partial r}(r_{\rm max}) =0\,.}
\label{BCdot}
\end{align}
Together with Eq.~(\ref{BCbin}), this condition ensures  that $|\mathbf{\hat{L}}|=1$ for all values of $r$ up to numerical errors \citep{2014MNRAS.441.1408T}.
\item
At the inner boundary, we expect Lense-Thirring precession to quickly align the disc with the BH spin and thus set 
\begin{align}
\mathbf{\hat L}(r_{\rm min}) =\mathbf{\hat {J}}\,.
\label{BCspin}
\end{align}
Note that Eq.~(\ref{BCspin}) corresponds to only two boundary conditions because Eq.~(\ref{BCdot}) already prescribes the magnitude of $\mathbf{\hat L}$. In practice, we impose ${\hat L}_x (r_{\rm min})={\hat L}_y (r_{\rm min})=0$ and let ${\hat L}_z(r_{\rm min})\ssim 1$ be determined by the solving algorithm.
\item
For a flat disc ($\psi=0$), the solution of Eq.~(\ref{tildealpha1}) reads \begin{equation}
\sigma(r)= \frac{2}{3}r^{-\beta} \left(1-\sqrt{\frac{r_{\rm ISCO}}{r}}\,\right)
\label{zerotorqueISCO}
\end{equation}
where the integration constant has been chosen to  impose a zero-torque boundary condition ($\sigma=0$) at the BH innermost stable circular orbit (ISCO); cf. \cite{2014MNRAS.441.1408T}. We assume $r_{\rm min}\gg r_{\rm ISCO}$ and obtain
 \begin{equation}
\sigma(r_{\rm min})= \frac{2}{3}r_{\rm min}^{-\beta}\label{sigmaBC}\,,
\end{equation}
which is our last boundary condition.
\end{itemize}
To ease convergence, we start from a flat disc without companion ($\theta=\kappa=0$) and progressively increase both $\theta$ and $\kappa$ providing the previous solution as initial guess to the BVP solver. 

{The misalignment angle of the outer disc $\theta$ enters the problem through the boundary condition of Eq.~(\ref{BCbin}).} Misalignments $\theta\leq 90^\circ$ ($\theta\geq 90^\circ$) correspond to co- (counter-) rotating discs. Equation~(\ref{coordass}) implies that a transformation $\theta\to \pi-\theta$ returns discs with identical shape but $\mathbf{\hat L}\cdot \mathbf{\hat J} \to - \mathbf{\hat J} \cdot \mathbf{\hat L} $. 

\section{Towards a consistent solution}
\label{sectowards}%

We now analyze the disc configuration using three approximations of increasing complexity. 

\subsection{The linear approximation}
\label{linear}

The warp amplitude $\psi= r|\partial\mathbf{\hat L}/\partial r|$ enters at $\mathcal{O}(\psi^2)$ in both the evolutionary equations (\ref{pringlemass}-\ref{pringlemom}) and the viscosity coefficients (\ref{alpha1og}-\ref{alpha2og}).  To linear order in $\psi$ one obtains
\begin{align}
&\frac{\partial \sigma}{\partial r} =   - \left( \beta  + \frac{1}{2}\right) \frac{\sigma}{r} + \frac{r^{-\beta -1}}{3}
\label{linear1}
\\
&\frac{\partial^2 \mathbf{\hat L}}{\partial r^2} =  \frac{\partial \mathbf{\hat L}}{\partial r} \bigg[-\frac{2 r^{-\beta-1} }{\zeta\,  \sigma} + \frac{3}{\zeta r}
- \left(\beta+\frac{3}{2}\right)\frac{1}{r} 
- \frac{1}{\sigma}\frac{\partial \sigma}{\partial r}  \bigg] 
\notag \\& 
-  r^{-\beta-3}\left(\mathbf{\hat J} \times \mathbf{\hat L}\right)
-\kappa   \,r^{-\beta+3/2}  \left(\mathbf{\hat L} \cdot \mathbf{\hat L}_\star\right) \left(\mathbf{\hat L} \times \mathbf{\hat L}_\star\right)\,,
\label{linear2}
\end{align}
and $\tilde\alpha_1(\alpha,\psi)=\tilde\alpha_2(\alpha,\psi)=1$.
This linear approximation is justified as long as the misalignment between the inner and the outer disc is small, $\theta\ll 1$. Accretion discs around spinning BH binaries in this regime have been  studied extensively by \cite{2009MNRAS.400..383M}. For $\kappa=0$, the solution can be written down in closed form using Bessel functions \citep{1996MNRAS.282..291S,2007MNRAS.381.1617M}. 

\subsection{Inconsistent non-linear treatment}
\label{inconsistent}

Next, one can inconsistently include terms of $\mathcal{O}(\psi^2)$ in the mass and momentum currents but neglect them when evaluating the viscosities. This approach has been pursued by \cite{2014MNRAS.441.1408T} using a numerical setup which is very similar to ours.
One needs to set $\tilde\alpha_i(\alpha,\psi)=1$ in Eqs.~(\ref{tildealpha1}-\ref{tildealpha2}) and solve
\begin{align}
&\frac{\partial \sigma}{\partial r} =   - \left( \beta  + \frac{1}{2}\right) \frac{\sigma}{r} - \frac{\zeta \sigma \psi^2}{3r} + \frac{r^{-\beta -1}}{3}  \,,
\label{notcons1}
\\
&\frac{\partial^2 \mathbf{\hat L}}{\partial r^2} =  \frac{\partial \mathbf{\hat L}}{\partial r} \bigg[\!-\frac{2 r^{-\beta-1}}{\zeta \sigma}  + \frac{3}{\zeta r}
 - \left(\beta+\frac{3}{2}\right)\frac{1}{r} 
- \frac{1}{\sigma}\frac{\partial \sigma}{\partial r}  \bigg] 
 \!-  \frac{\psi^2}{r^2}  \mathbf{\hat L}
\notag \\&- {r^{-\beta-3}} \left(\mathbf{\hat J} \times \mathbf{\hat L}\right)
-\kappa  { r^{-\beta+3/2}}  \left(\mathbf{\hat L} \cdot \mathbf{\hat L}_\star\right) \left(\mathbf{\hat L} \times \mathbf{\hat L}_\star\right).
\label{notcons2}
\end{align}

\subsection{Consistent non-linear treatment}
\label{iterative}

A consistent treatment requires taking into account all terms in Eqs.~(\ref{tildealpha1}-\ref{tildealpha2}). Unfortunately, the derivatives
\begin{align}
\frac{\partial \tilde\alpha_i(\alpha,\psi)}{\partial r}  = \frac{\partial \tilde\alpha_i(\alpha,\psi)}{\partial \psi} \left( \frac{\psi}{r}+ \frac{r^2}{\psi} \frac{\partial\mathbf{\hat L}}{\partial r }\cdot \frac{\partial^2 \mathbf{\hat L}}{\partial r^2}\right)
\end{align}
prevent writing down the expressions in normal form. \cite{2014MNRAS.441.1408T} opted for solving the full time-dependent dynamics until relaxation. Here we pursue a different approach.

We approximate the full solution using the following iterative scheme:
\begin{itemize}[leftmargin=0.6cm]
\item We first take $\tilde\alpha_i(\alpha,\psi)=1$ and solve the inconsistent problem reported in Eqs.~(\ref{notcons1}-\ref{notcons2}). 
\item The resulting warp profile $\psi(r)$ is used to evaluate the viscosities from Eq.~(\ref{alpha1og}-\ref{alpha2og}). We thus obtain numerical profiles $\alpha_1(\alpha,r)$ and $\alpha_2(\alpha,r)$.
\item These evaluations are used to approximate $\alpha_i(\alpha,\psi)$   in Eqs.~(\ref{tildealpha1}-\ref{tildealpha2}). We thus solve 
\begin{align}
&\frac{\partial \sigma}{\partial r} =   - \left( \beta  + \frac{1}{2}\right) \frac{\sigma}{r} - \frac{\zeta \sigma \psi^2}{3r} \frac{\tilde\alpha_2(\alpha,r)}{ \tilde\alpha_1(\alpha,r)}
\notag \\ &+ \frac{r^{-\beta -1}}{3 \tilde\alpha_1(\alpha,r)} - \frac{\sigma}{\tilde\alpha_1(\alpha,r)} \frac{\partial \tilde\alpha_1(\alpha,r)}{\partial r} \,,
\label{cons1}
\\
&\frac{\partial^2 \mathbf{\hat L}}{\partial r^2} =  \frac{\partial \mathbf{\hat L}}{\partial r} \bigg[-\frac{2 r^{-\beta-1} }{\zeta\, \tilde\alpha_2(\alpha,r) \sigma} + \frac{3}{\zeta r}\frac{ \tilde\alpha_1(\alpha,r)}{ \tilde\alpha_2(\alpha,r)} 
- \left(\beta+\frac{3}{2}\right)\frac{1}{r} 
\notag \\& 
- \frac{1}{\sigma}\frac{\partial \sigma}{\partial r} - \frac{1}{\tilde\alpha_2(\alpha,r)}\frac{\partial \tilde\alpha_2(\alpha,r)}{\partial r} \bigg] 
 -  \frac{\psi^2}{r^2}  \mathbf{\hat L}
 \notag \\&
-  \frac{r^{-\beta-3}}{\tilde\alpha_2(\alpha,r)} \left(\mathbf{\hat J} \times \mathbf{\hat L}\right)
-\kappa  \frac{ r^{-\beta+3/2}}{\tilde\alpha_2(\alpha,r)}  \left(\mathbf{\hat L} \cdot \mathbf{\hat L}_\star\right) \left(\mathbf{\hat L} \times \mathbf{\hat L}_\star\right)\,,
\label{cons2}
\end{align}
to obtain a new warp profile $\psi(r)$. 
\item The procedure is then iterated until convergence. Our convergence criterion is 
\begin{equation}
\max\Big(\max_r |\Delta \psi| ,\, \max_r |\Delta \tilde \alpha_1|,\, \max_r |\Delta \tilde \alpha_2|\Big)<10^{-3}\,,
\end{equation}
where the symbol $\Delta$ indicates the difference between two consecutive iterations. 
\end{itemize}

\subsection{Comparing the three approaches}

\begin{figure*}
\begin{tabular}{c@{\hskip 0.04in}c@{\hskip 0.04in}c}
\includegraphics[height=0.28\textwidth,page=1]{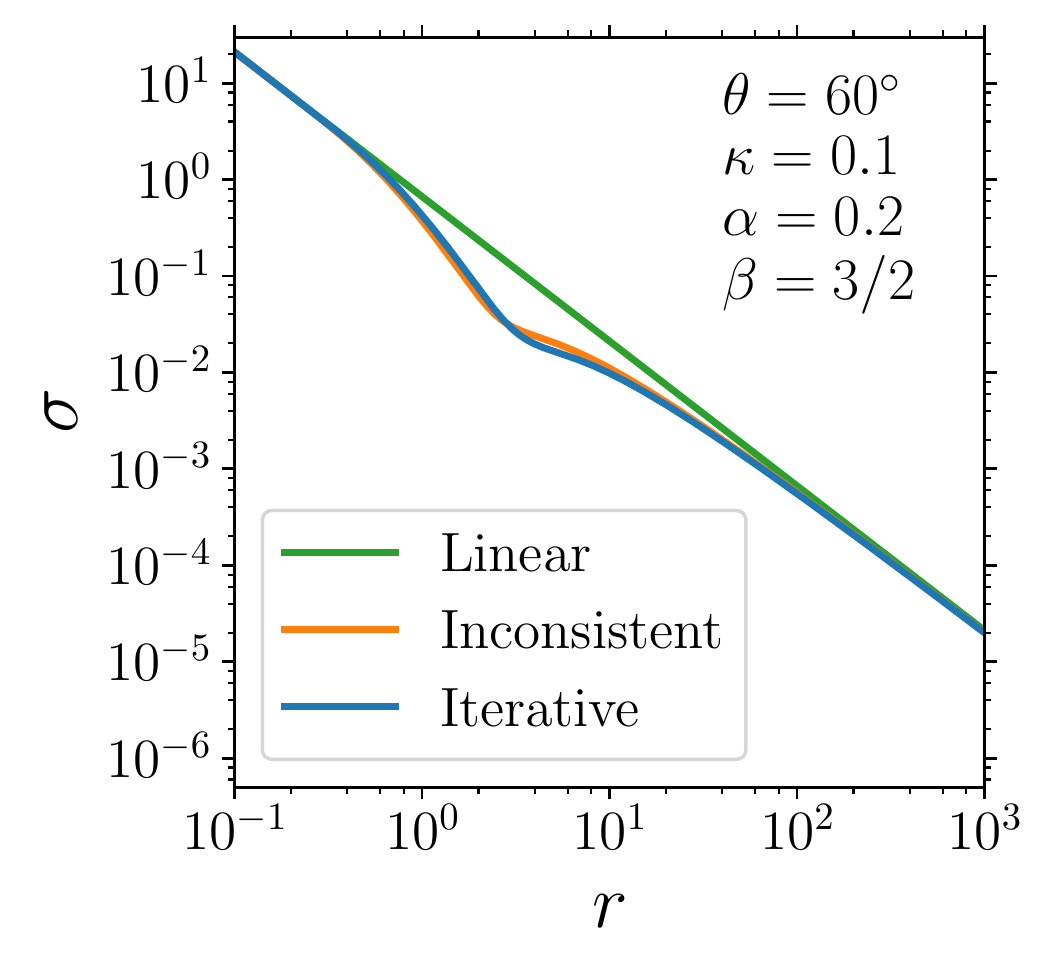}&
\includegraphics[height=0.28\textwidth,page=5]{differentapproximations.pdf}&
\includegraphics[height=0.28\textwidth,page=6]{differentapproximations.pdf}\\
\includegraphics[height=0.28\textwidth,page=2]{differentapproximations.pdf}&
\includegraphics[height=0.28\textwidth,page=3]{differentapproximations.pdf}&
\includegraphics[height=0.28\textwidth,page=4]{differentapproximations.pdf}\\
\includegraphics[height=0.28\textwidth,page=7]{differentapproximations.pdf}&
\includegraphics[height=0.28\textwidth,page=8]{differentapproximations.pdf}&
\includegraphics[height=0.28\textwidth,page=9]{differentapproximations.pdf}
\end{tabular}
\caption{{Disc profile for $\theta=60^\circ$, $\kappa=0.1$, $\alpha=0.2$, and $\beta=3/2$ under three approximations of increasing complexity: linear ({green}, Sec.~\ref{linear}), inconsistent (orange, Sec.~\ref{inconsistent}), and iterative ({blue}, Sec.~\ref{iterative}).} The disc presents two distinct regions, with a sharp transition located at the warp radius $r\ssim 1$. The inner region is aligned with the BH spin, i.e. $\hat L_x=0$, $\hat L_y=0$, and $\hat L_z\ssim 1$. The outer disc is aligned with the binary orbit, $\hat L_x=\sin\theta$, $\hat L_y=0$, and $\hat L_z= \cos\theta$. The rising of the warp $\psi$ at $r\ssim 1$ is paired to a sharp depletion of the surface density $\sigma$.  {The viscosities $\alpha_1$ and $\alpha_2$ are kept constant in the linear and inconsistent approaches;   orange and green curves thus coincide in the two lower left panels.} Our numerical grid ranges from $r=10^{-1}$ to $r=10^{4}$ and is here restricted to $r\leq10^{3}$ for illustrative purposes. %
}
\label{differentapproximations}
\end{figure*}

Figure~\ref{differentapproximations} shows a representative solution for $\theta=60^\circ$, $\kappa=0.1$, $\alpha=0.2$, and $\beta=3/2$. %
The disc is sharply divided between an inner region aligned with the BH spin and outer region aligned with the binary orbit. 

As expected, the transition happens at $r\ssim 1$, i.e. $R\ssim R_{\rm LT}$. As the warp amplitude increases, the surface density presents a pronounced drop. The two non-linear solutions capture a depletion in $\sigma$ of more than an order of magnitude compared to the flat-disc case where $\sigma\propto r^{-\beta}$.  This feature is absent in the linear disc profile since $\sigma\propto r^{-\beta}$ is the exact solution of Eq.~(\ref{linear1}). To the best of our knowledge, the relevance of this effect to the BH spin alignment problem has never been considered; we will discuss its impact in Sec.~\ref{sec:BHtorque}.

Because of some terms $\mathcal{O}(\psi^2)$ were neglected in Eq.~(\ref{tildealpha2}), in the linear approximation the magnitude $|\mathbf{\hat L}|$ differs from unity. Our algorithm returns values of $\mathbf|\mathbf{\hat L}|$ as large as $\ssim 6$ in the inner regions of the grid. Even for the $x$ and $y$ components of $\mathbf{\hat L}$ 
 that are supposed to be captured more accurately \citep{1996MNRAS.282..291S,2007MNRAS.381.1617M,2009MNRAS.400..383M}, we find that the linear approximation introduces errors of about $50\%$. %
For the two non-linear solutions, our numerical setup maintains the magnitude $|\mathbf{\hat L}|$ close to unity with an accuracy of $10^{-5}$ over the entire grid.  The largest numerical errors occur at $r\ssim r_{\rm min}$ because the boundary conditions (\ref{BCbin}-\ref{BCdot}) are imposed at $r_{\rm max}$.

For the consistent solution shown in Fig.~\ref{differentapproximations}, convergences was reached in 4 iterations. The viscosities $\alpha_1$ and $\alpha_2$ considerably depart from their unperturbed value at locations $R\ssim R_{\rm LT}$. However, their impact on the disc shape appears to be rather modest. The warp profile differ by only $\ssim10\%$ compared to the inconsistent case analyzed previously. More specifically, the iterative solution presents a smaller warp located at larger separations. However,  as clarified below, small differences and mismodeling in $\mathbf{L}$ at separations $R\lesssim R_{\rm LT}$ have a considerable impact on the spin-alignment time.

Unless explicitly mentioned, all disc profiles presented in this paper are computed using the iterative scheme described in Sec.~{\ref{iterative}}

\section{Spin alignment and critical obliquity}
\label{BPbin}
We now study the coupled evolution of the BH spin and its accretion disc, subject to the perturbation of a  binary companion orbiting at fixed orbital separation.

\subsection{Black-hole spin torque}
\label{sec:BHtorque}

The torque exerted by the disc onto the BH is given by the integral of the Lense-Thirring term in Eq.~(\ref{pringlemom}) along the disc profile, i.e. 
\begin{equation}
\frac{\mathrm{d}\mathbf{J}}{\mathrm{d}t} = - \int_{R_{\rm min}}^{R_{\rm max}} \frac{2G}{c^2}  \frac{\mathbf{J} \times \mathbf{ L}}{R^3}  2 \pi R \mathrm{d} R 
\end{equation}
where $R_{\rm min}$ and $R_{\rm max}$ mark the extent of our numerical grid.
The evolution of the misalignment angle $\theta$ is given by:
\begin{align}
\frac{\mathrm{d}\cos\theta}{\mathrm{d}t} 
=\frac{\mathrm{d} \mathbf{\hat J}}{\mathrm{d} t} \cdot \mathbf{ \hat{L}_\star} 
= - \frac{1}{t_{\rm align}} \int_{r_{\rm min}}^{r_{\rm max}} \!\! (\mathbf{\hat J} \times \mathbf{\hat L}) \cdot \mathbf{ \hat L_\star}  \frac{\sigma} {r^{3/2}}  \,\,\mathrm{d}r
\label{dcosthetadt}
\end{align}
where %
\begin{align}
t_{\rm align} 
&= \frac{1}{\dot M}\sqrt{\frac{c}{G}M\chi \frac{\alpha\nu_0}{\zeta}}\,.%
\end{align}
From Eqs~(\ref{eddfraction}) and (\ref{fullnu0}) one gets 
\begin{align}
t_{\rm align} 
&=
6.2 \times 10^6 
 \left(\!\frac{\chi}{0.5} \right)^{2/3}
\left(\frac{H_0/R_0}{0.002} \right)^{2/3}  
\notag \\&\times
 \left(\frac{f}{0.1} \right)^{-1}
\left(\frac{\alpha}{0.2} \right)^{1/3}
\left[\frac{\zeta}{1/(2\!\times\! 0.2^2)} \right]^{-2/3} {\rm yr}\,,
\label{talign}
\end{align}
 which agrees with earlier derivations by \cite{1998ApJ...506L..97N} and \cite{2013MNRAS.429L..30L}.  

The  mass of the BH increases on a timescale
$t_{\rm acc}\simeq M/\dot M$. One obtains
\begin{equation}
\frac{t_{\rm align}}{t_{\rm acc}}\simeq \alpha^{5/3}\chi^{2/3}\,\left(\frac{H_0}{R_0}\right)^{2/3}
\label{taligntacc}
\end{equation}
The disc, on the other hand, readjust its shape due to the external torque on the viscous timescale $t_{\nu 2}\ssim R_0^2/\alpha_2 \nu_0$, which yields %
\begin{equation}
\frac{t_{\rm align}}{t_{\nu2}}\simeq \frac{c^3}{G\dot M} \alpha^{-4/3}\chi^{-1/3}\,\left(\frac{H_0}{R_0}\right)^{14/3}\,.
\label{taligntnu}
\end{equation}
Equations~(\ref{taligntacc}) and (\ref{taligntnu}) were obtained using Eq.~(\ref{nu0HR}), approximating $\zeta\simeq 1/\alpha^2$, assuming $R_0=R_{\rm LT}$, and omitting factors of order unity.

For a representative AGN disc with ${H_0}/{R_0}\ssim 10^{-3}$ and $\alpha\ssim 0.1$ %
 feeding a BH of  $M\ssim 10^7 M_\odot$  at (a fraction of) the Eddington rate, one obtains:
\begin{equation}
t_{\nu 2}\ll t_{\rm align}\ll t_{\rm acc}\,.
\end{equation}

The first inequality describes the canonical Bardeen-Petterson effect \citep{1975ApJ...195L..65B,1976IAUS...73..225R}. The inner regions of the disc  quickly align with the BH spin on the timescale $t_{\nu 2}$. On the longer time $t_{\rm align}$, the outer disc pulls the BH towards a complete aligned configuration. The spin alignment process can thus be studied in a quasi-adiabatic fashion assuming a sequence of steady-state disc solutions,  justifying the assumptions made in Sec.~\ref{warpeddiscs}. 
The second inequality implies that the change in mass of the BH can be safely neglected during the entire evolution.

In the bottom-right panel of Fig.~\ref{differentapproximations} we plot the integrand of Eq.~(\ref{dcosthetadt}), thus illustrating how each gas ring contributes to the evolution of the misalignment $\theta$. In our coordinate system one has $-(\mathbf{\hat J} \times \mathbf{\hat L}) \cdot \mathbf{ \hat L_\star} = \hat L_y \sin\theta$. Moreover, the component $L_y$ vanishes at both the inner and the outer boundary; cf. Eq.~(\ref{coordass}). Only the central region where the disc is warped contributes meaningfully to the alignment process. The effect of the warp is counterbalanced by the depletion of the surface density $\sigma$ at those same locations. More specifically, the innermost regions of the disc where $L_y\lesssim 0$ tend to increase the BH misalignment. The disc annuli closer to the warp radius where $L_y\gtrsim 0$, however, provide larger contributions to the torque and ultimately drive the system towards $\theta\to0 $ (or equivalently  $\mathbf{\hat J}\to\mathbf{\hat L}_\star$). 

While the disc shape has been previously solved at the non-linear level \citep{2014MNRAS.441.1408T}, to the best of our knowledge these solutions have never been used to compute the alignment torque. For the linear, inconsistent, and iterative case shown in Fig.~\ref{differentapproximations} we obtain $t_{\rm align}\times d\cos\theta/dt= 0.34, 0.19$, and 0.21, respectively. Therefore, employing the linear warp approximation to study the spin alignment problem results in an underestimate of the alignment time of about $50\%$. This is because the linear case does not capture the depletion of the surface density at the warp radius and therefore overestimates the alignment torque. The iterative treatment of the viscosities presented in Sec.~\ref{iterative} introduces a $\ssim 10\%$ correction.

Initial misalignments larger than $\pi/2$ deserve a separate discussion. A transformation $\theta\to \pi-\theta$ corresponds to $L_y\to L_y$ and $\sin\theta\to \sin\theta$, and, therefore, does change the sign of the derivative ${\rm d}\cos\theta/{\rm d}t$. Even for initially counter-rotating discs, the dynamics always tend to co-align the disc and the BH \citep{1996MNRAS.282..291S}. %
As first pointed out by \cite{2005MNRAS.363...49K}, counter-alignment is a possible outcome only for discs with small enough angular momentum. In this study, we anchor the disc at $\mathbf{\hat L}_\star$ at the outer edge of our numerical grid, thus assuming that the angular momentum of the disc is much larger than the BH spin. %

To investigate the validity of this assumption, let us consider the angular momentum of a Keplerian disc $L_\mathrm{disc} \simeq M_\mathrm{disc} \sqrt{G M R_\mathrm{out}} $, where $ M_\mathrm{disc}$ is the disc mass and $R_\mathrm{out}$ is the disc extent. The former can be written as $ M_\mathrm{disc} \simeq \dot{M} t_\nu  = f M t_\nu / t_\mathrm{edd}$ where  $t_\nu = R_{\rm out}^2/\nu$  is the viscous time. Using the \citet{1973A&A....24..337S} prescription one obtains
\begin{align}
\frac{L_\mathrm{disc}}{J} & = \frac{f}{\chi} \frac{c}{G M t_\mathrm{edd}} \frac{R_\mathrm{out}^2}{\alpha} \left(\frac{H}{R} \right)^{-2}
\notag \\& \simeq 85 \left( \frac{f}{0.1} \right) \left( \frac{\chi}{0.5} \right)^{-1} 
\left( \frac{M}{10^7 M_\odot} \right)^{-1} \left( \frac{R_\mathrm{out}}{0.05 \mathrm{pc}} \right)^{2} \notag \\& \times  \left( \frac{\alpha}{0.02} \right)^{-1} \left( \frac{H/R}{0.002} \right)^{-2},
\end{align}
where here $H/R$ is the aspect ratio at  $R_{\rm out}$.  At least at the beginning of the phase in which the disc drives the inspiral ($R_\star\ssim 0.05$ pc, cf. Sec.~\ref{secjoint}), one has $L_{\rm disc}\gg J$ which justifies our boundary conditions.

\subsection{The shape of the disc}
\label{shapeofthedisc}

The shape of an accretion disc surrounding a BH in a binary system depends on four parameters: 
\begin{enumerate}[leftmargin=0.8cm]
\item the outer misalignment angle $\theta$, 
\item the contribution of the companion $\kappa$, 
\item the kinematic viscosity coefficient $\alpha$,
\item and the viscosity spectral index $\beta$. 
\end{enumerate}
We now systematically address the impact of these quantities.

\begin{figure*}
\flushleft
\hspace{2.8cm}\includegraphics[scale=0.45]{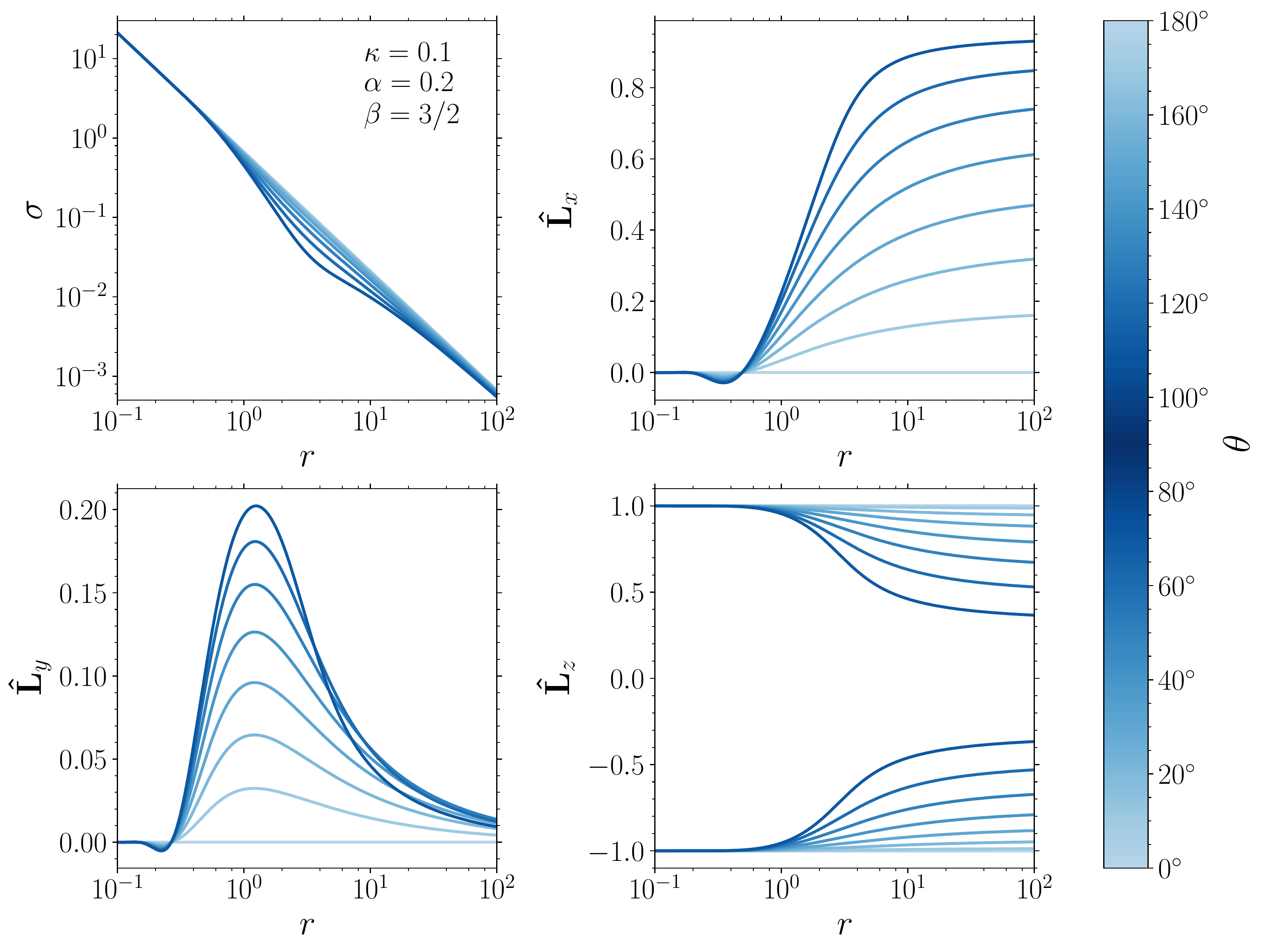}

\caption{Sequence of discs with different obliquities  $\theta=0^\circ, 10^\circ, 20^\circ, 30 ^\circ, 40^\circ, 50^\circ, 60^\circ,70^\circ, 110^\circ, 120^\circ, 130^\circ, 140^\circ, 150^\circ, 160^\circ, 170^\circ, 180^\circ$ (light to dark) and fixed values of  $\kappa=0.1$, $\alpha=0.2$, and $\beta=3/2$. The outer misalignment  $\theta$ sets the boundary condition for $\mathbf{\hat L}$ and determines the depletion of $\sigma$ at the warp radius. The symmetry $\theta\to\pi-\theta$ leaves  $\sigma$, $\mathbf{\hat L}_x$, $\mathbf{\hat L}_y$ unchanged (hence two profiles overlaps for each visible curve) and transforms $\mathbf{\hat L}_z \to -\mathbf{\hat L}_z $ (hence the two sets of curves in the bottom-right panel).}
\label{varytheta}

\vspace{0.5cm}

\hspace{2.8cm}\includegraphics[scale=0.45]{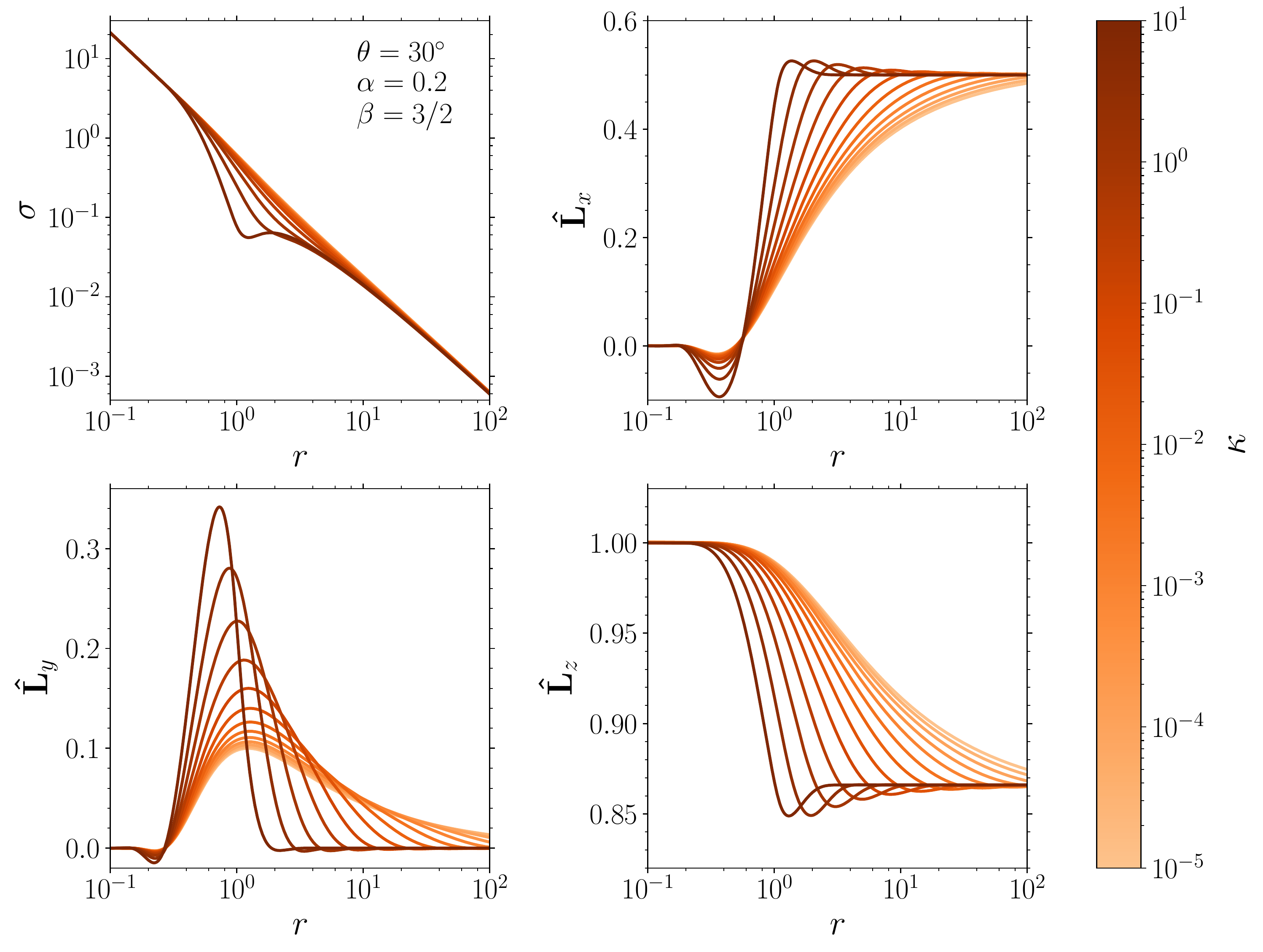}
\caption{Sequence of discs with different companion parameter $\log_{10}\kappa=-5,  -4.5, -4, -3.5, -3, -2.5, -2, -1.5, -1, -0.5, 0, 0.5, 1$ (light to dark) and fixed values of $\theta=30^\circ$, $\alpha=0.2$, and $\beta=3/2$. The parameter $\kappa$ determines the location of the warp radius. In particular, larger (smaller) values of $\kappa$ correspond to cases where the companion torque is more (less) relevant and present discs with a smaller (larger) warp radius.} 
\label{varykappa}
\end{figure*}

\begin{figure*}
\flushleft
\hspace{2.8cm}\includegraphics[scale=0.45]{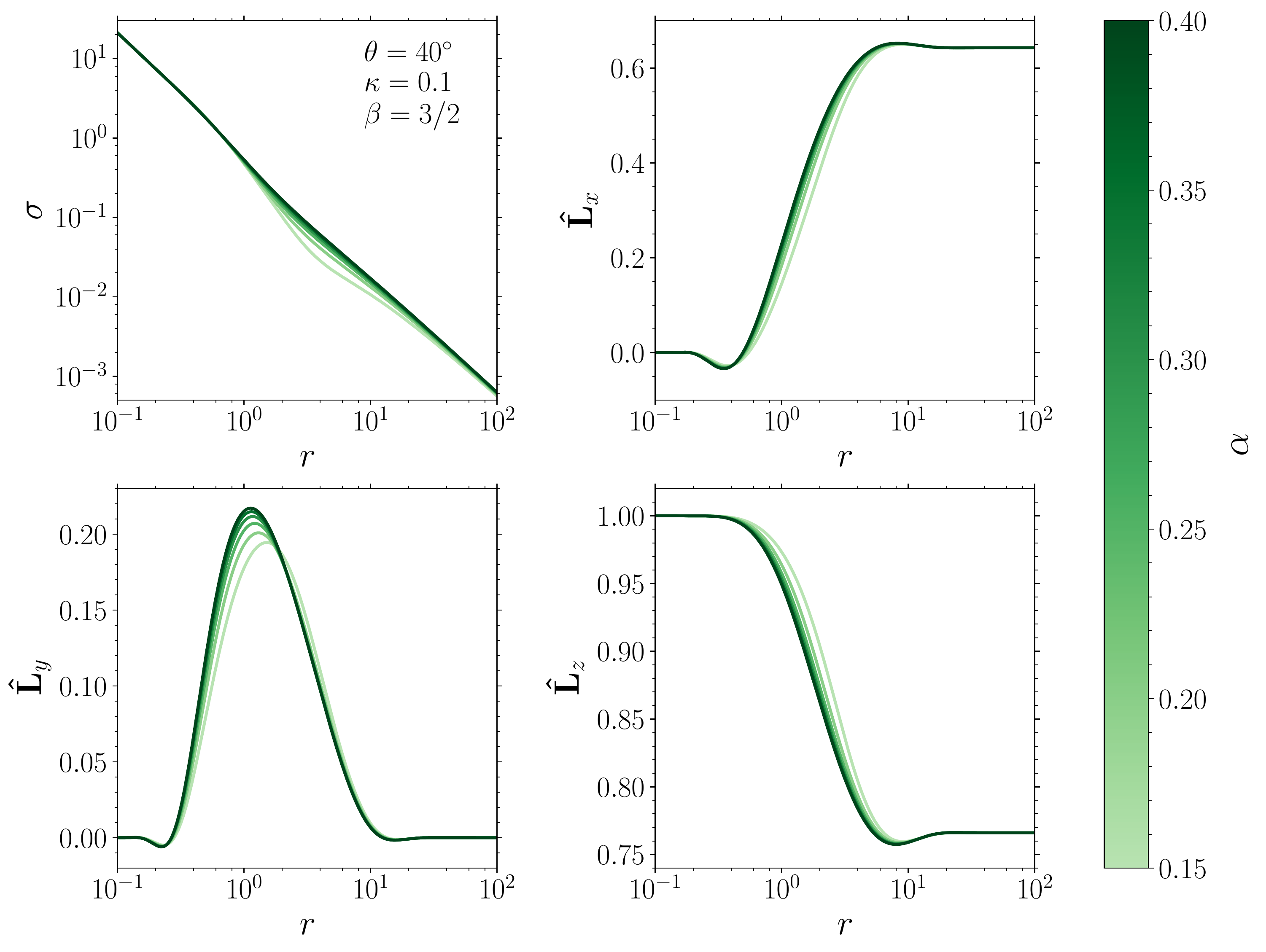}
\caption{
Sequence of discs with different kinematic viscosity $\alpha=0.15,0.2,0.25,0.3,0.35,0.4$ (light to dark) and fixed values of $\theta=40^\circ$, $\kappa=0.1$, and $\beta=3/2$. \emph{If solutions can be found}, the coefficient $\alpha$ appears to have a marginal effect on the shape of the disc in dimensionless units.
}
\label{varyalpha}

\vspace{0.5cm}

\hspace{2.8cm}\includegraphics[scale=0.45]{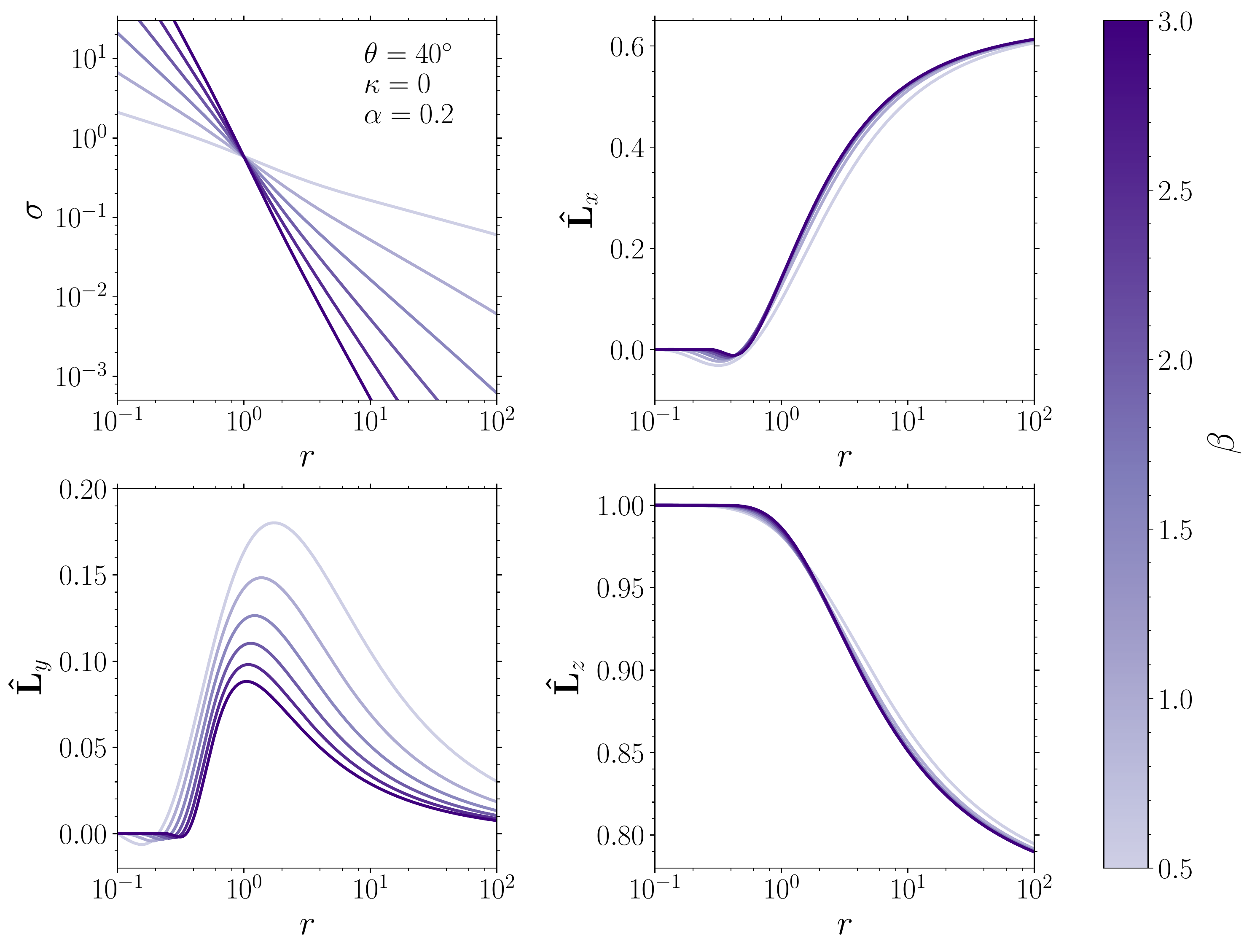}
\caption{Sequence of discs with different viscosity slope $\beta=0.5,1,1.5,2,2.5,3$ (light to dark) and fixed values of $\theta=40^\circ$, $\kappa=0$, and $\alpha=0.2$. The parameter $\beta$ sets the slope of the surface density $\sigma$. The impact of $\beta$ on the disc angular momentum is largely restricted to the $y$ component, while  $\mathbf{\hat L}_x$ and $\mathbf{\hat L}_z$ are almost unchanged.} 
\label{varybeta}
\end{figure*}

Figure~\ref{varytheta} shows a sequence of discs with progressively higher obliquity $\theta=0^\circ,\dots,70^\circ$ and $110^\circ,\dots,180^\circ$. We fix $\kappa=0.1$, $\alpha=0.2$, and $\beta=3/2$. %
The angle $\theta$ sets the depletion of the surface density $\sigma$. More inclined discs present  sharper transitions between the inner and the outer regions, and consequently a lower value of $\sigma$ at the warp radius. The location of the warp radius  radius itself is largely independent of $\theta$. The disc is sensibly warped only for a small portion of its radial extent: for the parameters chosen in Fig.~\ref{varytheta}, the warp concentrates between $r\sim 0.2$ and $r\sim 10$. Figure~\ref{varytheta} also illustrates that the symmetry $\theta\to \pi-\theta$ reverses the sign of the component of $\mathbf{\hat L}$ parallel to the BH spin, while leaving $L_x$ and $L_y$ unchanged.

Figure~\ref{varykappa} examines the impact of the BH companion, which is encoded in the parameter $\kappa$. We vary $\kappa=10^{-5}, \dots,10^{1}$ and fix $\theta=30^\circ$, $\alpha=0.2$,  $\beta=3/2$. As  $\kappa$ departs significantly from $0$, the transitions between the inner and the outer disc becomes sharper. The more relevant the companion, the more the warp radius moves inwards toward the central BH. The presence of the companion partially counterbalances the Lense-Thirring torque and allows gas rings to stay misaligned closer to the accreting object. This effect was first pointed out by   \cite{2013ApJ...774...43M}.

In Figure~\ref{varyalpha}  we vary the Shakura-Sunyaev coefficient $\alpha=0.15,\dots,0.4$ for a series of discs with $\theta=40^\circ$, $\kappa=0.1$ and $\beta=3/2$. When solutions can be found, $\alpha$ has a subdominant impact and leaves the shape of the disc in dimensionless units almost unchanged. In general, smaller values of $\alpha$ correspond to slightly sharper warp profiles with lower surface density. It should be noted, however, that the $\alpha$ coefficient sets the physical scale of the Lense-Thirring radius [Eq.~(\ref{kappavalue})], as well as the disc mass [Eq.~(\ref{dimlessscaling})] and the alignment timescale [Eq.~(\ref{talign})], and it is thus a crucial parameter once scaling to physical units. Furthermore, $\alpha$ has the crucial role of determining when solutions do or do not  exist. This point is explored in Sec.~\ref{nonexistencesec}. For the case shown in Fig.~\ref{varyalpha}, solutions cannot be found for $\alpha\lesssim 0.08$. 

Finally, in Fig.~\ref{varybeta} we study the relevance of the parameter $\beta=0.5,\dots, 3$ for discs with $\theta=40^\circ$, $\kappa=0$, and $\alpha=0.2$. The isothermal case studied so far corresponds to $\beta=3/2$. The coefficient $\beta$ sets the slope of the viscosities which, to linear order, is equal to the opposite of the spectral index of the surface density: $\sigma\propto r^{-\beta} + \mathcal{O}(\psi^2)$. Smaller (larger) values of $\beta$ therefore corresponds to discs with shallower (steeper) mass density profiles. By definition, all curves have the same surface density at $r=1$; see Eqs.~(\ref{nu1}-\ref{nu2}). The behaviour of the angular momentum is less intuitive: $\beta$ appears to affect only the projection $(\mathbf{\hat J} \times \mathbf{\hat L}) \cdot \mathbf{ \hat L_\star}\propto L_y$. Notably, this is the only component that enters the alignment process, cf. Eq.~(\ref{dcosthetadt}). Profiles with smaller (larger) values of $\beta$ corresponds to disc profiles which are more (less) bended in the $y$ direction. %

\begin{figure}\centering
\includegraphics[width=0.9\columnwidth]{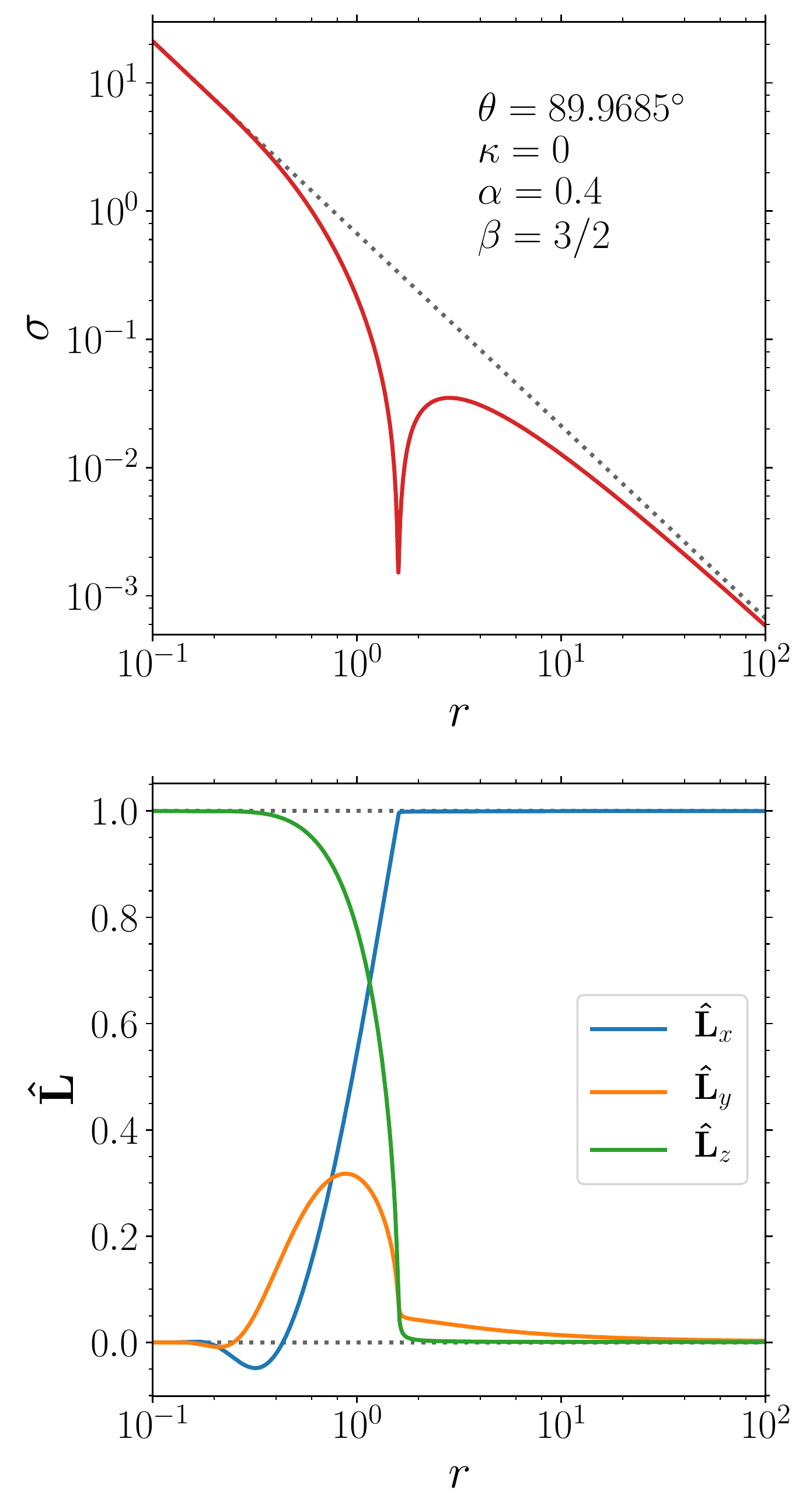}
\caption{Surface density (top panel) and angular momentum (bottom panel) of a warped disc near criticality. We fix $\kappa=0$, $\alpha=0.4$, $\beta=3/2$ and progressively increase the obliquity $\theta$ in steps of $5\times 10^{-4}$. Here we report the last converged solution, obtained for  $\theta=89.9685^\circ$. The disc is essentially broken into two disjoint regions: an inner disc aligned with $\mathbf{\hat z}$ and an outer disc aligned with $\mathbf{\hat x}$. Numerical errors for this profile are $\big||\mathbf{\hat L}| -1\big|\lesssim 1.5\times 10^{-6}$ over the entire grid. Dotted lines show  flat discs with the same value of $\beta$.}
\label{critical}
\end{figure}

\subsection{The critical obliquity}
\label{nonexistencesec}

For some regions of the parameter space, our BVP algorithm does not converge. This same behaviour was found  by \cite{2014MNRAS.441.1408T} with different integration methods. In general, physical configurations cease to exist for values of $\theta$ close to $90^\circ$, large values of $\kappa$, and small values of $\alpha$. 

As highlighted  in Sec.~\ref{shapeofthedisc}, large values of $\theta$ and $\kappa$ correspond to steeper and steeper warp profiles. Eventually, the transition between the inner and the outer disc becomes too sharp to be resolved.  
A near-critical case is shown in Fig.~\ref{critical}. In practice, these configurations correspond to two completely disjoint discs: an inner disc aligned to the BH spin and an outer disc with misalignment $\theta$.

\begin{figure}\centering
\includegraphics[width=\columnwidth]{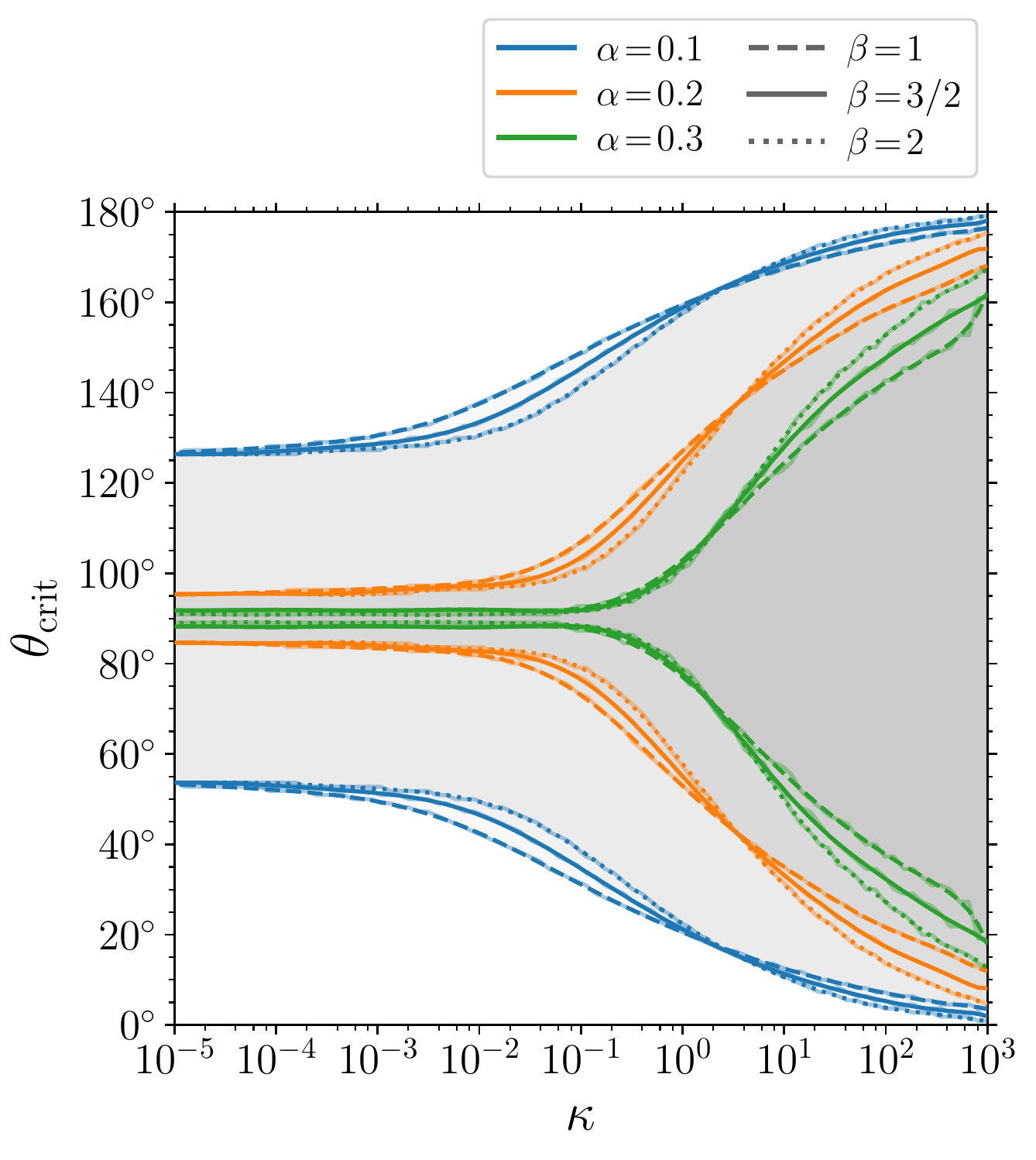}
\caption{Regions of the parameter space where physical solutions can or cannot be identified. In particular, we show the critical obliquity $\theta_{\rm crit}$ as a function of $\kappa$ ($x$-axis), $\alpha$ (line colors), and $\beta$ (line styles). Solutions are found only in the white/lighter areas ranging from $\theta=0^\circ$ and $180^\circ$ until each of the curves. Transparent curves underneath mark the results of our integrations; smoother curves on top show polynomial fits. %
}
\label{goodbadline}
\end{figure}

Figure~\ref{goodbadline} shows the allowed region in the $\theta$-$\kappa$ parameter space. In particular, for each $\kappa$ we compute the critical obliquity $\theta_{\rm crit}\leq \pi/2$ below (above) which solutions can (cannot) be found. The situation is reversed for $\theta\geq \pi/2$: solutions are (not) found only for values of $\theta$ greater (smaller) than the critical obliquity $\theta_{\rm crit}$. There appear to be two different regimes. For $\kappa \gtrsim 1$, the critical obliquity changes rather sharply until most of the parameter space is excluded. For $\kappa\lesssim 1$, on the other hand, $\theta_{\rm crit}$  asymptotes to a constant value. 

\begin{figure}\centering
\includegraphics[width=\columnwidth]{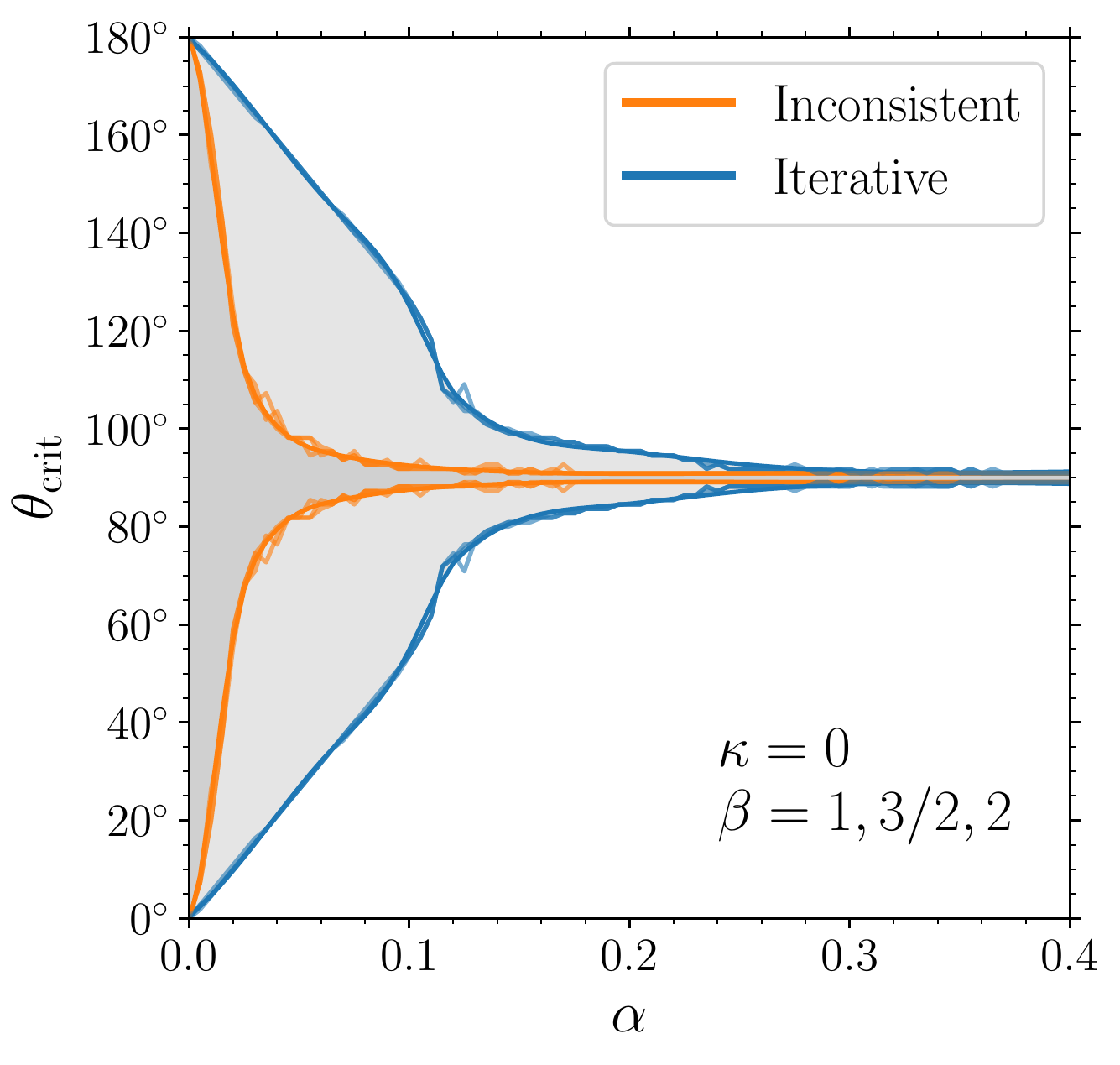}
\caption{Critical obliquity $\theta_{\rm crit}$ as a function of the viscosity coefficient $\alpha$ for the isolated case $\kappa=0$. Results are shown for three values of $\beta=1,3/2,2$ and appear indistinguishable. Solutions are found only in the white/lighter areas below each of the curves. Transparent curves underneath mark the results of our integrations; smoother curves on top show polynomial fits.}

\label{nobinexist}
\end{figure}

Figure~\ref{nobinexist} shows the critical obliquity for $\kappa=0$ (i.e. the asymptote in Fig. \ref{goodbadline}) as a function of $\alpha$ and $\beta$.  The region where solutions are not found is largely independent of $\beta$ but  increases dramatically for $\alpha\lesssim 0.1$. In this regime, the non-linear warp theory of \citet{2013MNRAS.433.2403O} predicts negative viscosities for moderate warp values $\psi\lesssim 1$ (cf.~\citealt{2018MNRAS.476.1519D}). Our BVP solver is unable to find consistent solution whenever this condition is approached. For comparison, Fig.~\ref{nobinexist} also shows the critical misalignment for the inconsistent case described in Sec.~\ref{inconsistent}, where the viscosity coefficients $\alpha_1$ and $\alpha_2$ are not allowed to vary with $r$. In this case, the BVP converges over a much larger region $\alpha\gtrsim 0.01$. %
In any case, we are never able to solve a BVP for exactly  orthogonal discs $\theta=90^\circ$ (cf. \citealt{2014MNRAS.441.1408T}).

The viable region of the parameter space is only mildly influenced by the slope of the surface density. Larger $\beta$ correspond to slightly larger (smaller) critical obliquity for small (large) values of $\kappa$; cf. Fig.~\ref{goodbadline}.

During the lifetime of a BH binary, disc migration tends to increase $\kappa$ while the Lense-Thirring torque tends to decrease $\theta$. Physical BHs will trace paths starting from the top-left  toward the bottom-right corner of Fig.~\ref{goodbadline}. Depending on their trajectories in this plane, sources might become critical in finite time. We will study this issue in Sec.~\ref{spinwithinsp}.

\subsection{Spin alignment}

The evolution of the spin orientation $\theta(t)$ can be found integrating the projected torque reported in  Eq.~(\ref{dcosthetadt}). Both $\theta$ and $\kappa$ are function of time and need to be integrated together. For illustrative purposes, we first integrate ${\rm d}\theta/ {\rm d}t$ keeping $\kappa$ fixed and postpone the complete problem to  the next section.

\begin{figure}\centering
$\,$\\$\,$\\
\includegraphics[width=\columnwidth]{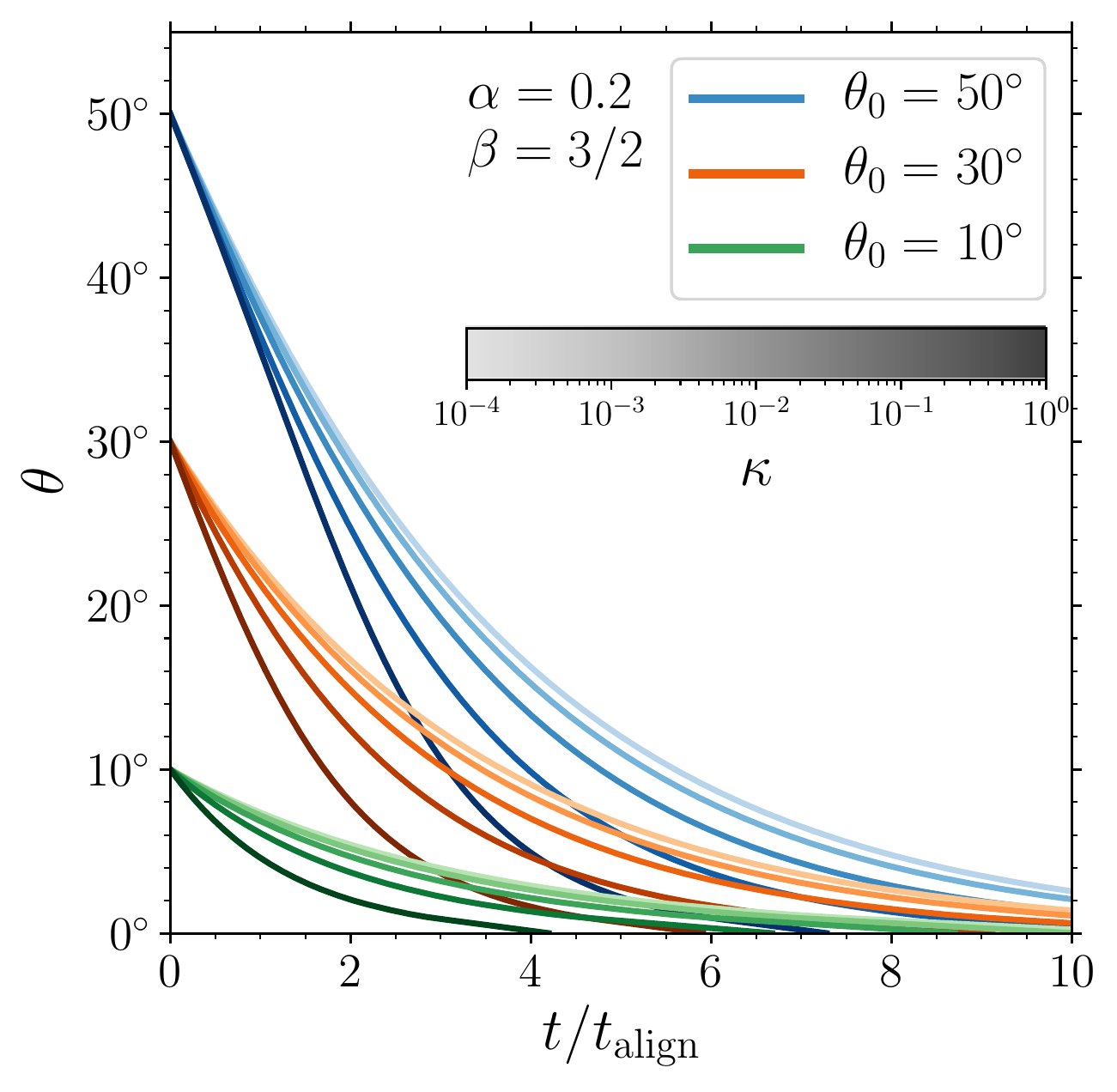}
\caption{Evolution of the disc-spin misalignment $\theta$ as a function of time. We assume $\alpha=0.2$, $\beta=3/2$, different initial misalignments $\theta_0=10^\circ, 30^\circ, 50^\circ$ (colors) and different values of $\kappa=10^{-4},10^{-3},10^{-2},10^{-1},1$ (lighter to darker). In this figure we artificially keep $\kappa$ constant. For this set of parameters, the smallest critical obliquity (corresponding to the largest value $\kappa=1$) is $\theta_{\rm crit}\ssim 58^\circ$. %
}
\label{thetaoft}
\end{figure}

Figure~\ref{thetaoft} shows the evolution $\theta(t)$ for a set of discs with $\alpha=0.2$, $\beta=3/2$, and $\kappa=10^{-4},\dots, 1$. The behavior resembles that of an exponential $\theta(t)\simeq \theta_0 \exp(-t/t_{\rm align})$. Indeed, the analytical solution of \citet{1996MNRAS.282..291S} and \citealt{2007MNRAS.381.1617M} valid in the linear regime shows that an exponential is the solution in the limit of small angles (more accurately, it is $\sin \theta$ that decreases exponentially). %
 The parameter $\kappa$ introduces variations of order unity, with larger $\kappa$ corresponding to faster spin alignment  \citep{2013ApJ...774...43M}. For instance, starting from $\theta_0=50^\circ$, systems with $\kappa=1$ ($\kappa= 0$) are found at $\theta\ssim2^\circ$ ($\theta\ssim 10^\circ$) after $t\ssim 5\times t_{\rm align}$.

As identified previously, the values of $\alpha$ and $\beta$ have a minor impact on the dimensionless misalignment process. The viscosity coefficient $\alpha$, however, enters the time scale $t_{\rm align}\propto  \alpha^{1/3} \zeta^{-2/3}\sim \alpha^{5/3}$ [Eq.~(\ref{talign})]: the lower the viscosity, the faster spins  align.

\begin{figure}\centering
\includegraphics[width=\columnwidth]{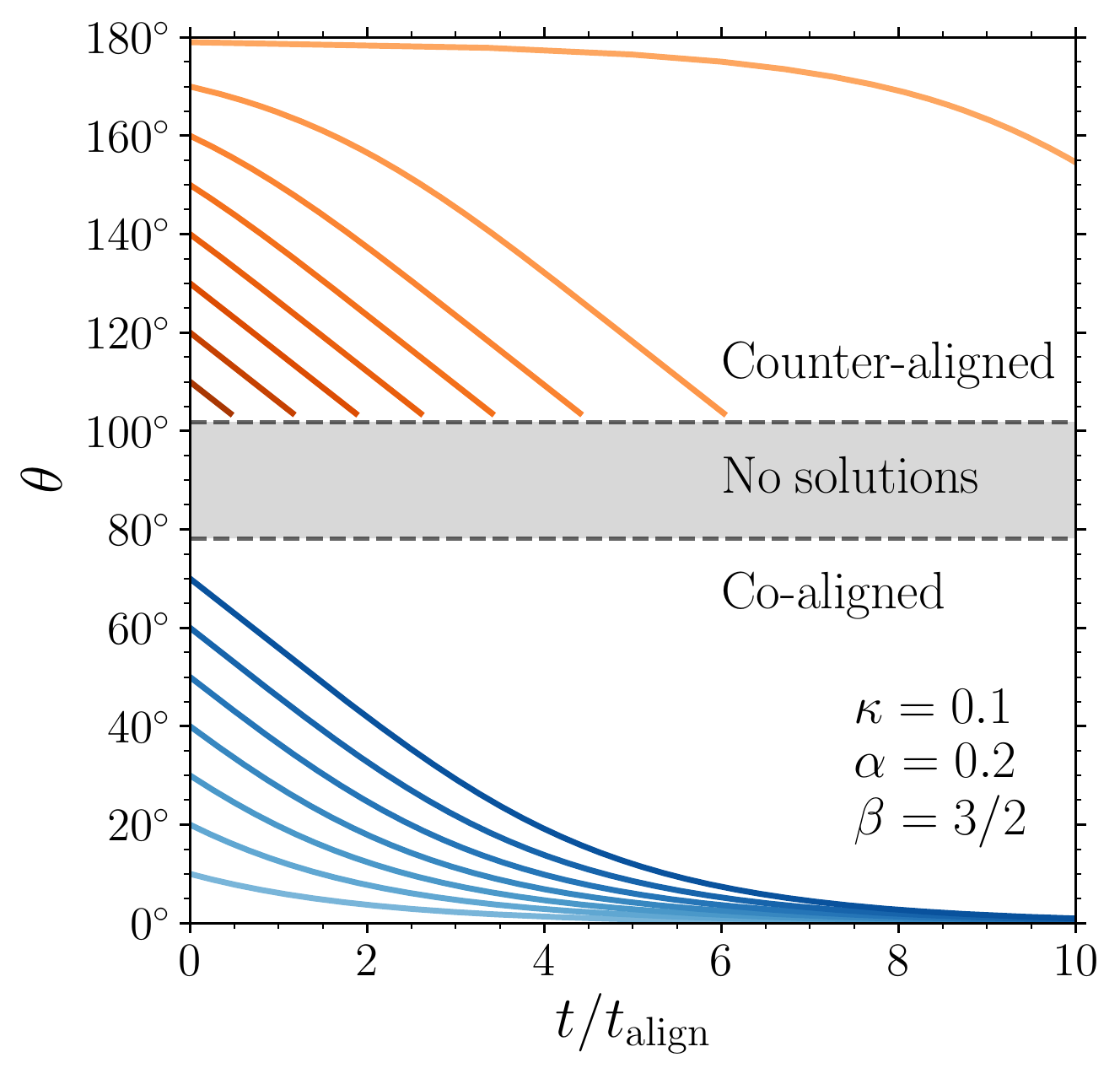}
\caption{Evolution of the spin misalignment $\theta$ with time for initially co-rotating (blue, $\theta<90^\circ$) and counter-rotating (orange, $\theta>90^\circ$) inner discs. The dashed lines mark the critical obliquity $\theta_{\rm crit}$. Disc solutions cannot be found in the grey area. This figure is produced assuming $\alpha=0.2$, $\beta=3/2$, $\kappa=0.1$, and neglecting the time evolution of $\kappa$.}
\label{counterdown}
\end{figure}

Irrespectively of the initial obliquity, the dynamics always tend to co-align the spin and the disc, i.e. the angle $\theta$ decreases with time. %
The expected phenomenology is summarized in Fig.~\ref{counterdown}. For discs initially co-aligned with the BH spin ($\theta<\pi/2$) the evolution can take place only if the initial angle $\theta_0$ is below the critical obliquity $\theta_{\rm crit}$. In this case, the system aligns on a timescale given by Eq.~(\ref{talign}). For  initially counter-aligned discs ($\theta>\pi/2$), the system will reach the critical obliquity on this same timescale. The fate of the system in this scenario is unclear and might be related to the disc breaking studied by \cite{2012MNRAS.421.1201N,2013MNRAS.434.1946N,2015MNRAS.448.1526N}.

\section{Joint inspiral and alignment}
\label{secjoint}

We now investigate the importance of the binary inspiral on the disc surrounding each BH. As the orbital separation $R_\star$ decreases, the parameter  $\kappa$ increases, thus moving the warp radius inwards and speeding up the alignment. At the same time, larger values of $\kappa$ shrink the region where physical solutions are present.

\subsection{Inspiral parametrization}\label{sec:evol_theta_inspiral}

The development of a complete model of supermassive BH migration in binaries, possibly including information from large-scale cosmological simulations, is outside the scope of this paper and is postponed to a future publication. For now, we implement simple prescriptions that capture only the key features in a parametrized fashion.

We assume that all of the mass from the circumbinary disc is accreted by either of the two BHs, thus neglecting potential pile-up at the edge of the cavity carved up by the binary.  This assumption is supported by some \citep{2014ApJ...783..134F,2015ApJ...807..131S}, but not all \citep{2013MNRAS.436.2997D,2016MNRAS.460.1243R}, recent contributions on the topic. The accretion rate of the circumbinary disc is thus given by the sum of the individual contribution $\dot M + \dot M_\star$. 
 Hydrodynamical simulations \citep{2014ApJ...783..134F} (but see also \citealt{2015MNRAS.452.3085Y}) suggest that the ratio between the accretion rates of the two BHs scales as
\begin{equation}
\label{diffacc}
\frac{\dot M}{\dot M_\star} = \frac{M_\star}{M}\,,
\end{equation}
which implies \emph{differential accretion} \citep{2015MNRAS.451.3941G}: the smaller (larger) BH accretes more (less) mass from the circumbinary. 
If $f$ is the Eddington fraction of the disc surrounding the BH of mass $M$ from Eq~(\ref{eddfraction}), the Eddington fraction of the circumbinary disc is given by\footnote{\cite{2015MNRAS.451.3941G} make use of a different notation where $f$ is Eddington fraction of the circumbinary disc, while here $f$ refers to the disc of the aligning BH.}
\begin{equation}
f_{M+M_\star} 
= t_{\rm Edd} \frac{\dot M + \dot M_\star}{M+M_\star} 
=  f \frac{M}{M_\star}\,.
\label{fcbin}
\end{equation}

We assume that the time  a BH binary spends a given separation $R_\star$ is given by a power law with spectral index $\gamma$ scaled at values $t_{\rm b}$ and $R_{\rm b}$, i.e.
\begin{equation}
\label{tinspiral}
t_{\rm inspiral} = \frac{t_{\rm b}}{f_{M+M_\star}}   \left(\frac{R_\star}{R_{\rm b}}\right)^\gamma \,. \\
\end{equation}
The  model developed by \cite{2015MNRAS.451.3941G} based on Type-II planetary migration predicts $\gamma$ between 0 (if the binary dominates) and $3/2$ (if the disc dominates) (see also \citealt{1995MNRAS.277..758S,2013ApJ...774..144R,2015MNRAS.448.3603D}). \cite{2009ApJ...700.1952H} reports $1/2\leq \gamma\leq 11/4$ depending on various  assumptions on the disc structure.
As for the normalization,  previous works by \cite{2003MNRAS.339..937G,2005ApJ...630..152E,2009ApJ...700.1952H,2017MNRAS.469.4258T,2017MNRAS.464.3131K,2019MNRAS.482.4383F} reported inspiral timescales of few to tens of Myr from separations $R_{\rm b}\sim 0.05$ pc. For moderate Eddington fractions $f_{M+M_\star}\sim 0.1$, this corresponds to $t_{\rm b}\sim10^6$ yr. For more context, let us note that \cite{2012ApJ...749..118S} found larger values $t_{\rm b} = t_{\rm Edd}/0.8\simeq 5\times 10^8$ yr, while \cite{2020ApJ...889..114M} found that the binary gains angular momentum from the disc instead of losing it.

The coupled problem of inspiral and alignment  consists of the following set of ODEs
\begin{align}
\frac{{\rm d} R_{\star}}{{\rm d} t} &=  - \frac{R_{\star}}{t_{\rm inspiral}(R_\star)}
\label{dRstar}
\\
 \frac{{\rm d}\cos \theta}{{\rm d} t} &= \frac{\mathrm{d}\, \mathbf{\hat J}}{\mathrm{d} t} \! \cdot\! \mathbf{ \hat{L}_\star}\;(\theta,R_\star)\,,
 \label{dtheta}
\end{align}
with  initial conditions $\theta=\theta_0$ and $R_\star=R_{\star 0}$.

The right-hand side of Eq.~(\ref{dtheta}) depends on $R_{\star}$ only through $\kappa$. One can rewrite Eqs.~(\ref{dRstar}-\ref{dtheta}) as 
\begin{align}
\label{dcosthetadlogkappa}
 \frac{{\rm d}\cos \theta}{{\rm d} \ln\kappa} &=  - \omega\, {\kappa}^{-\gamma/3}
\int_{r_{\rm min}}^{r_{\rm max}} (\mathbf{\hat J} \times \mathbf{\hat L}) \cdot \mathbf{ \hat L_\star}  \frac{\sigma} {r^{3/2}}  \,\,\mathrm{d}r 
 \end{align}
 where we introduced the dimensionless quantity 
 \begin{align}
 \omega = \frac{\kappa_{\rm b}^{\gamma/3}}{3 f_{M+M_\star}}\frac{t_{\rm b}}{t_{\rm align}} \,,
 \end{align}
 and $\kappa_b$ is the value of $\kappa$ at $R_{\rm b}$.
We integrate Eq.~(\ref{dcosthetadlogkappa}) numerically by interpolating a grid of precomputed disc profiles. In this simplified model, the corresponding evolution of the separation and the elapsed time can be derived analytically. One gets
\begin{align}
&R_{\star} = R_{b}\left(\frac{\kappa}{\kappa_{\rm b}}\right)^{-1/3}\,,
\\
&t= \begin{cases} \displaystyle
\frac{t_{\rm b}}{ \gamma \, f_{M+M_\star}} \left[\left(\frac{R_{\star 0}}{R_{\rm b}}\right)^\gamma-\left(\frac{R_{\star}}{R_{\rm b}}\right)^\gamma\right] \quad &{\rm if}\quad \gamma\neq 0\,,
\\
\displaystyle
\frac{t_{\rm b}}{f_{M+M_\star}} \ln \left(\frac{R_{\star 0}}{R_\star}\right) \quad &{\rm if}\quad \gamma= 0\,.
\end{cases}
\end{align}

Intuitively, the evolution $\theta(t)$ is set by two ingredients: the integral in Eq.~(\ref{dcosthetadlogkappa}) contains information on the shape of the disc, while the parameter $\omega$ encodes the relative importance of the inspiral and alignment processes. With the prescriptions of Eqs. (\ref{kappavalue}), (\ref{talign}), and (\ref{fcbin}) one obtains 
\begin{align}
\label{longomega}
{\omega} &	{\simeq \left(0.54 \times 10^{0.55 \gamma} \right)
\left(\frac{M}{10^7 M_\odot}\right)^{-1+2\gamma/3}
\left(\frac{\chi}{0.5}\right)^{2(\gamma-1)/3}}
\notag\\
&{\times \left(\frac{M_\star}{10^7 M_\odot}\right)^{1+\gamma/3}
\left(\frac{R_{\rm b}}{0.05 {\rm pc}}\right)^{-\gamma}
\left(\frac{t_{\rm b}}{10^6 {\rm yr}}\right)}
\notag\\
&
{\times\left(\frac{H_0/R_0}{0.002}\right)^{-2(\gamma+1/3)}
 \left(\frac{\alpha}{0.2}\right)^{-\gamma-1/3}
\left[\frac{\zeta}{1/(2\!\times\! 0.2^2)} \right]^{2/3-\gamma}\,.}
\!\!\!\,
\end{align}
Although here we have assumed simple prescriptions, we stress that our model is rather flexible: more accurate circumbinary disc physics (for instance where the aspect ratio is allowed to depend on other quantities) will still result in Eq.~(\ref{dcosthetadlogkappa}) but with a different expression for~$\omega$.

\subsection{Spin evolution during the inspiral}
\label{spinwithinsp}

\begin{figure}\centering
\includegraphics[width=0.41\textwidth,page=1]{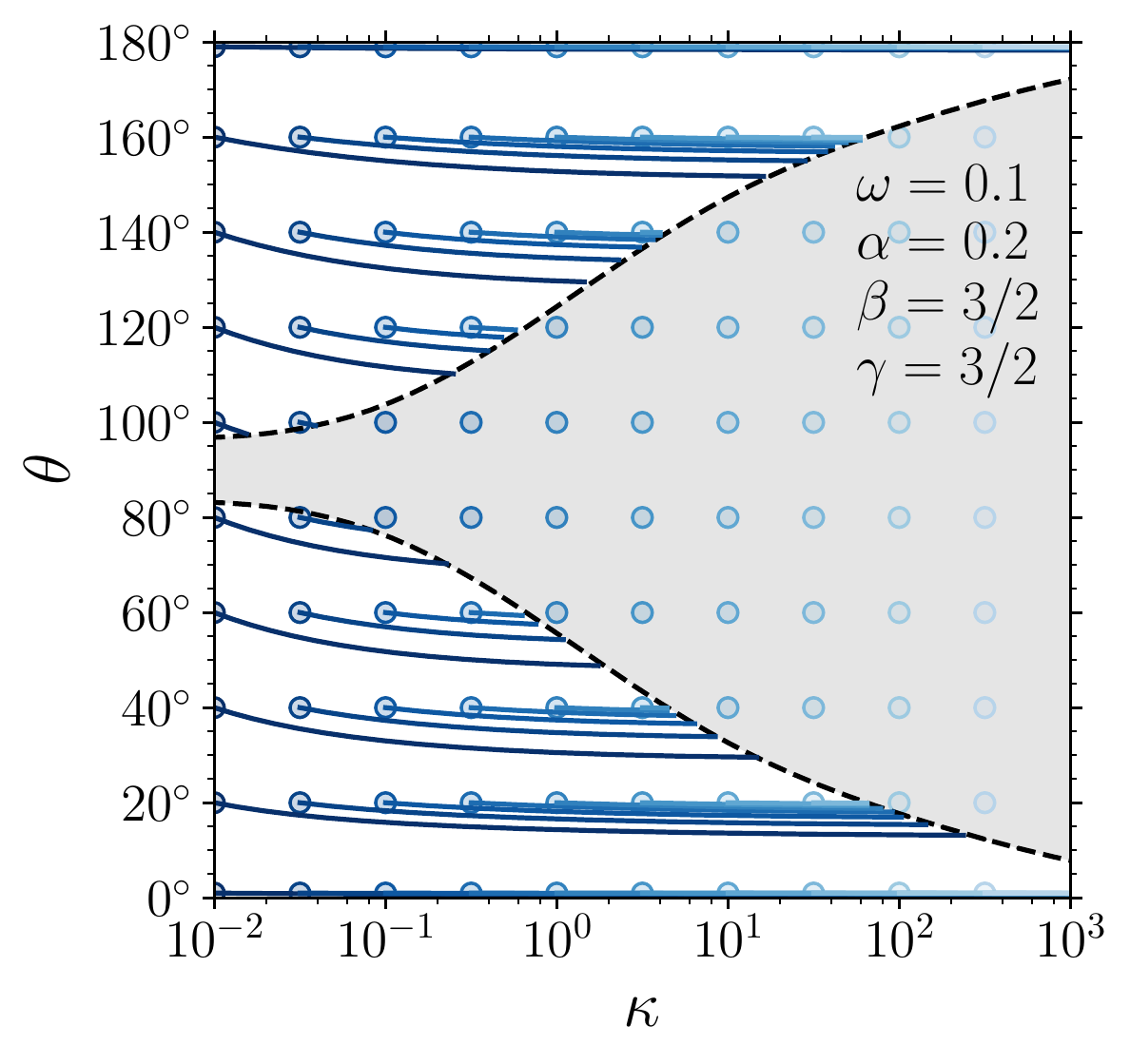}\\
\includegraphics[width=0.41\textwidth,page=2]{inspiralmany}\\
\includegraphics[width=0.41\textwidth,page=3]{inspiralmany}
\caption{Coupled evolution of the spin angle $\theta$ and the binary parameter $\kappa$ during the inspiral. Circles mark the initial conditions. The evolutionary tracks are indicated with blue curves: as the inspiral proceeds, the angle  $\theta$ decreases and the companion parameter $\kappa$ increases. Dashed black curves mark the critical obliquity $\theta_{\rm crit}$, beyond which solutions cannot be found (gray shaded areas). Top, middle, and bottom panel assume $\omega=$ 0.1, 1, and 10, respectively. All panels  are produced with $\alpha=0.2$, $\beta=3/2$, and $\gamma=3.2$.}
\label{inspiralmany}
\end{figure}

\begin{figure*}
\includegraphics[width=0.99\textwidth]{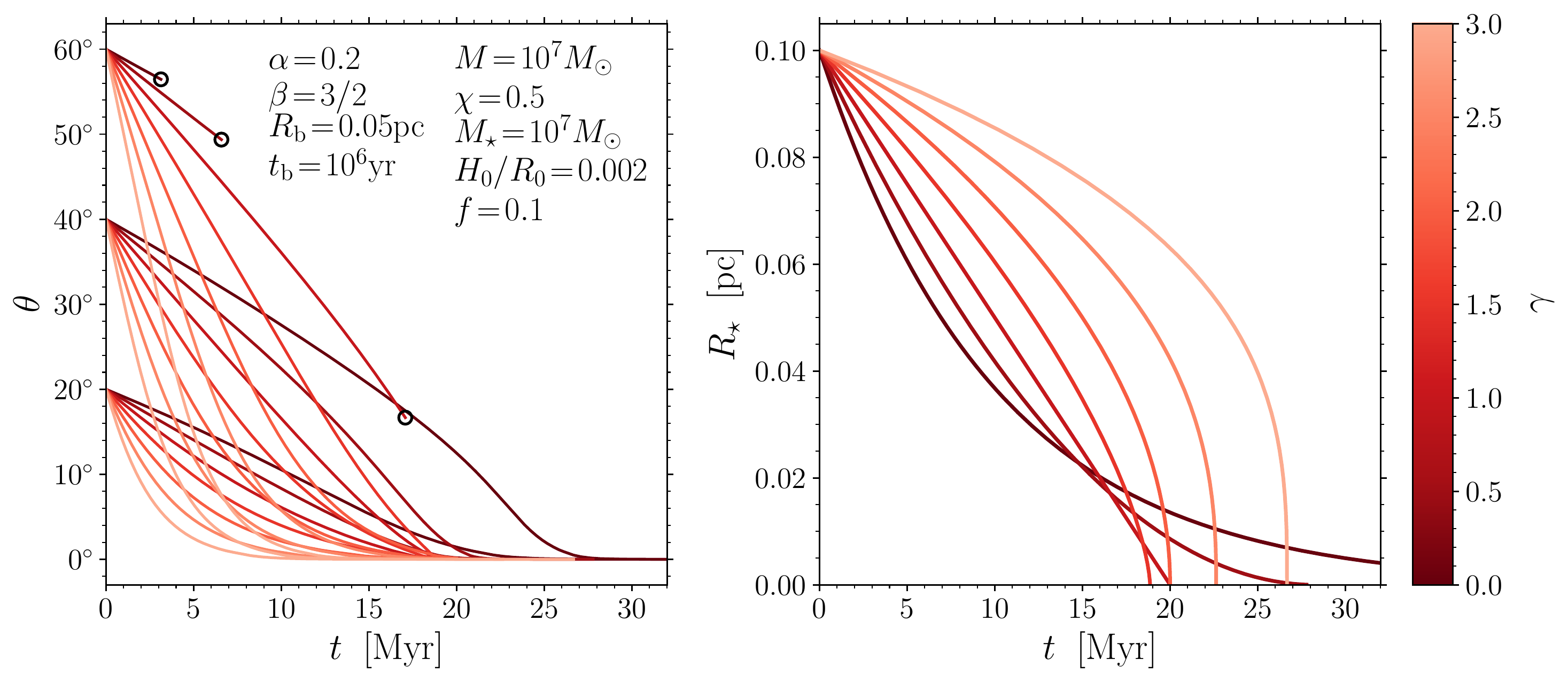}
\caption{Evolution of  spin angle $\theta$ (left-hand panel) and orbital separation $R_{\star}$ (right-hand panel) as a function of time. We present a sequence of integrations characterized by different values of the inspiral-time spectral index $\gamma=0, 0.5, 1, 1.5, 2, 2.5, 3$ (lighter to darker).
We assume $\alpha=0.2$, $\beta=3/2$, $M=M_{\star}=10^7 M_\odot$, $\chi=0.5$, $H_0/R_0=0.002$, $f=0.1$, $R_{\rm b}=0.05$ pc, and $t_b=10^6$ yr. {Integrations are initialized at $R_{\star 0}=0.1$ pc (corresponding to  $\kappa_0\simeq 0.32$) and three misalignment angles $\theta_0=20^\circ, 40^\circ, 60^\circ$}. Black circles in the left panel correspond to  critical configurations $\theta=\theta_{\rm crit}$ where disc solutions cease to exist. %
}
\label{inspiralgamma}
\end{figure*}

Figure~\ref{inspiralmany} shows some evolutionary tracks in the $(\theta\!-\!\kappa)$ plane for $\alpha=0.2$, $\beta=\gamma=3/2$, and $\omega=0.1,1,10$. Evolutions proceed from the top-left to the bottom-right region of the plots: as  binaries inspiral toward merger,  spins align ($\theta$ decreases) and companions become more important ($\kappa$ increases).  

Crucially, there are two possible outcomes. Some of the sources reach full alignment $\theta\ssim 0^\circ$ already for moderate values of $\kappa$. On the other hand, other systems meet the critical obliquity $\theta_{\rm crit}$ at some point during the inspiral. In our model, this happens for all systems with $\theta>90^\circ$ and some of the systems with $\theta<90^\circ$. The fate of these binaries needs to be further investigated: it is unclear if/how the disc can sustain the alignment process beyond criticality.  

The parameter $\omega\propto t_{\rm b}/t_{\rm align}$ determines the decrease in $\theta$ for a given increment in $\kappa$. Larger (smaller) values of $\omega$ correspond to shorter (longer) alignment times compared to the inspiral time. The evolution of $\theta(\kappa)$ is thus steeper (flatter) and less (more) systems reach the breaking point $\theta_{\rm crit}$. In particular, one has $\theta(t)\simeq$ constant for $\omega\to 0$ (implying that most discs reach the critical obliquity) and $\kappa(t)\simeq$ constant for $\omega\to \infty$ (implying that most systems fully align). For $\omega\sim 1$ (middle panel of Fig~\ref{inspiralmany}), inspiral and alignment roughly balance each other and the outcome of each configuration is the result of the interplay between the two processes.

\begin{figure}
\includegraphics[width=0.99\columnwidth]{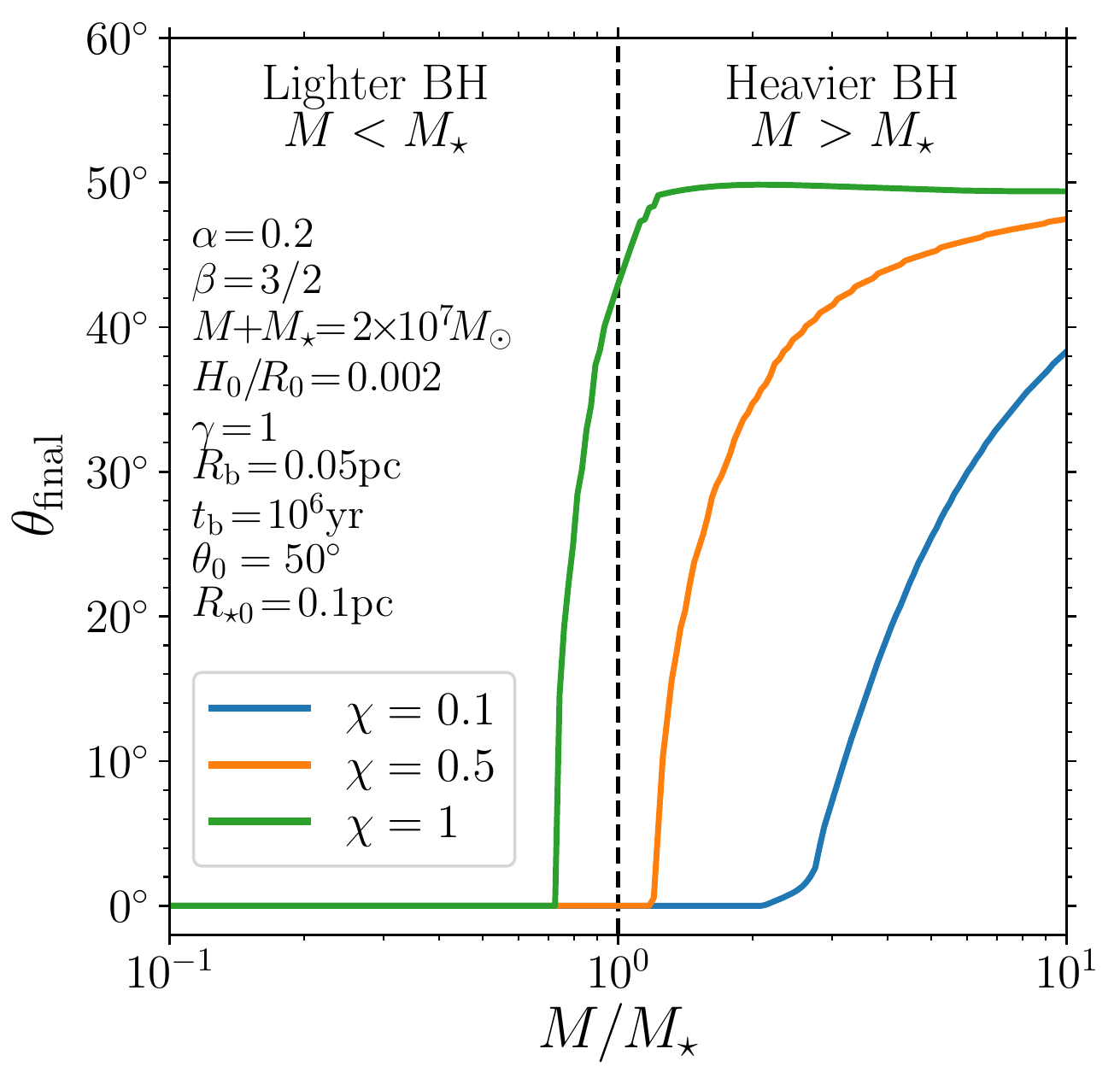}
\caption{Final angles $\theta_{\rm final}$ as a function of the binary mass ratio $M/M_\star$. {We set $\alpha=0.2$, $\beta=3/2$, $\gamma=1$, $H_0/R_0=0.002$, $R_{\rm b}=0.05$ pc, $t_b=10^6$ yr; integrations are initialized at $R_{\star 0}=0.1$ pc and $\theta_0=50^\circ$.}  In particular, we present three sequences of integrations where the total mass of the binary is kept fixed to $M+M_{\star}=2\times 10^7 M_\odot$, and the mass ratio of the companion varies between $M_\star=M/10$ and $M_\star=10 M$. The reported value $\theta_{\rm final}$ refers to the spin alignment of the BH with mass M: this is either the secondary (left region of the plot) or the primary (right region) component of the BH binary.  Blue, orange, and green lines show results obtained for dimensionless spin $\chi=0.1, 0.5,$ and 1, respectively. %
 }
\label{MoverMstar}
\end{figure}

Figure~\ref{inspiralgamma} illustrates the role of the inspiral-time slope $\gamma$. {We integrate Eqs.~(\ref{dRstar}-\ref{dtheta}) from $R_{\star 0}=0.1$ pc and three angles $\theta_0=20^\circ, 40^\circ, 60^\circ$; the other parameters are set to: $\alpha=0.2$, $\beta=3/2$, $M=M_{\star}=10^7 M_\odot$, $\chi=0.5$, $H_0/R_0=0.002$, $f=0.1$, $R_{\rm b}=0.05$ pc,  $t_b=10^6$ yr. } The index $\gamma$ is varied from 0 to 3, thus including all the values predicted by \cite{2009ApJ...700.1952H} and \cite{2015MNRAS.451.3941G}. {For these integrations, the parameter $\omega\propto 10^{(0.55\gamma)}\simeq 3.5^\gamma$ ranges from $\ssim 0.6$ (for $\gamma=0$) to $\ssim 13.8$ (for $\gamma=3$). BH spins in systems characterized by larger (smaller) values of $\gamma$ tend to align faster (slower).
Some binaries reach full alignment ($\theta=0$), while others reach the critical condition (black circles). The configurations which are more likely to become critical are those with $\gamma\ssim 0$.}

{Within the assumption of this study, we predict that viscous accretion can escort BH spins only to some final angle $\theta_{\rm final}$:} this is either $\ssim 0$ or the critical value where solutions cease to be present. In Fig.~\ref{MoverMstar} we explore the dependence of $\theta_{\rm final}$ on the binary mass ratio $M/M_{\star}$ for a sequence of binaries with fixed total mass $M+M_{\star}=2\times 10^7 M_\odot$. {The other parameters are set to: $\alpha=0.2$, $\beta=3/2$, $\chi=0.5$, $H_0/R_0=0.002$, $\gamma=1$, $R_{\rm b}=0.05$ pc,  $t_b=10^6$ yr (note that the Eddington fraction $f$ is irrelevant in this case, because it does not enter either $\kappa$ or $\omega$)}.  Let us stress that, in this paper, the mass of the aligning BH is denoted with $M$, while the symbol $M_{\star}$ indicates the mass of the companion. Therefore, the left region of  Fig.~\ref{inspiralgamma} where $M<M_{\star}$ refers to the spin alignment of the secondary, lighter component of the BH binary. Conversely, in the right region one has $M>M_{\star}$ and the reported misalignments refer to the primary, heavier binary member. 

We find that secondary BHs tend to align quickly while primaries remain close to their initial orientations $\theta_0$ (which is set to $60^\circ$ in Fig.~\ref{MoverMstar}). This is a direct consequence of the {differential accretion} prescription introduced in Eq.~(\ref{diffacc}): if the companion is sufficiently light, accretion on the primary BH  is heavily suppressed which, in turn, suppresses the alignment.  As expected, BH with larger spins $\chi$ have larger alignment time and thus are less likely to reach $\theta\sim 0$.
The results of Fig.~\ref{MoverMstar} confirms previous findings by some of the authors \citep{2015MNRAS.451.3941G}, albeit with an important caveat: systems which do not align reach the critical obliquity. The fate of those discs and BHs remains unclear.

\section{Conclusions}
\label{secconcl}

In this paper we presented a critical re-investigation of the Bardeen-Petterson effect  in supermassive BH binaries \citep{1975ApJ...195L..65B}. The alignment of BH spins with the angular momentum of their accretion discs is determined by the general-relativistic Lense-Thirring torque integrated over the disc profile. The largest contribution comes from gas rings at the location of the disc where viscous and relativistic drags balance each other (the ``warp radius'').

\subsection{Key results}

We showed that the commonly employed linear approximation to the warp dynamics underestimates the alignment time by up to $50\%$. We presented a new iterative scheme to capture the non-linear behavior of the fluid viscosities at all orders in the warp amplitude. We predict a strong depletion of the mass surface density at the warp radius, which diminishes the effectiveness of the Bardeen-Petterson effect resulting in longer alignment times. 

The formalism here developed takes into account the perturbation to the circum-BH disc induced by the binary companion, encoded in a single dimensionless parameter $\kappa$ [Eq.~(\ref{kappavalue})]. The torque from the BH companion decreases the warp radius, allowing material to stay misaligned closer to the accreting object and thus speeding up the alignment. We also presented a simplified treatment of the joint inspiral-alignment problem, and showed that this can also be parametrized by a single dimensionless quantity $\omega$ [Eq.~(\ref{longomega})].

Companion torque and warp non-linearities determine, together, whether a solution to the stationary, one-dimensional accretion-disc equations can be found. If the misalignment angle of the outer disc is larger than a ``critical obliquity'', viable solutions cease to exists.  
This specific feature of the Bardeen-Petterson effect was first pointed out by \cite{2014MNRAS.441.1408T} and is here explored in greater detail. 
We find that generic systems might reach the critical obliquity on a timescale of $\ssim 10^6$ yr.
The issue is more severe and impacts a larger portion of the parameter space if $\kappa$ is large (because the perturbation due to the companion grows) and/or  $\alpha$ is small (because warps deviate more strongly from their linear regime). Moreover, we find that all configurations that are initially counter-aligned (i.e. $\theta>\pi/2$) reach the critical obliquity in some finite time.

The key message of our paper is that surface-density depletion, companion perturbation, warp non-linearities, and critical obliquity must all be taken into account to predict the alignment between the BH spins and their accretion discs. In particular, we predict that all systems reach one of two possible endpoints: either complete alignment or a critical configuration.

\subsection{Importance of the disc structure}

The fate of the disc and the binary at the critical obliquity is unclear and constitutes an important area of future research. Our speculation is that the disc might break into two disjoint sections (an inner disc aligned with the BH spin and an outer disc beyond critical obliquity) which are not in viscous contact with each other (cf. \citealt{2012MNRAS.421.1201N,2013MNRAS.434.1946N,2015MNRAS.448.1526N}). This claim needs to be backed up by hydrodynamical simulations.

Our disc profiles now need to be put into proper context. For instance, in our simplified model we assumed that the aspect ratio of the circum-BH disc   $H_0/R_0$ has a constant value of $O(10^{-3})$. In reality, this parameter is set by the disc microphysics and depends on the central mass $M$, the $\alpha$ coefficient, and the accretion rate  $\dot M$ \citep{1973A&A....24..337S,2009ApJ...700.1952H}. Because $\kappa$ depends on $H/R$ to a very steep power  [Eq.~(\ref{kappavalue})], a proper disc model is crucial  to correctly determine the perturbation due to the companion  and thus understand which regions of the parameter space fall beyond the critical obliquity. %

A self-consistent disc profile is also important to properly initialize the Bardeen-Petterson integrations.
A value for the largest extent of the circumbinary disc is provided by its fragmentation radius. This is can be estimated from  \citeauthor{1964ApJ...139.1217T}'s (\citeyear{1964ApJ...139.1217T}) criterion
\begin{equation}
Q\equiv \frac{c_s \Omega}{\pi  G \Sigma} = 1
\end{equation}
where $\Omega = \sqrt{G(M+M_\star)/R_\star^3}$ is the Keplerian angular velocity of the circumbinary disc and $c_s=H \Omega$ is the speed of sound in the thin-disc approximation. Using $\dot M = 3 \pi \nu\Sigma$ and $\nu = \alpha c_s H$ one finds 
\begin{align}
&R_{\rm frag}\simeq \left(\frac{H}{R}\right)^2 \left(3 \alpha \frac{t_{\rm Edd}}{f_{M+M_\star}}\right)^{2/3} [G(M+M_\star)]^{1/3}
\notag \\
&\simeq
0.035  
 \left(\!\frac{M+M_\star}{2\!\times \!10^7 M_\odot}\!\right)^{1/3}
\!\! \left(\!\frac{H/R}{0.002}\!\right)^{2}
\!\! \left(\!\frac{f_{M+M_\star}}{0.1}\!\right)^{-2/3}
\!\!\left(\!\frac{\alpha}{0.2}\!\right)^{2/3}
\!\!{\rm pc}\,,
\end{align}
where here $H/R$  is the aspect ratio of the circumbinary disc at $R_{\rm frag}$. Disc fragmentation is a further ingredient\footnote{The toy integrations shown in Fig.~\ref{inspiralgamma} are initialized with orbital separation $R_{\star 0}\gtrsim R_{\rm frag}$ to better showcase the resulting phenomenology.} that determines the region of the parameter space that is forbidden by the critical obliquity and will need to be investigated carefully.

\subsection{Further caveats}

We assumed that mass and spin magnitude of the BH do not change because of the accreted material. This is a well justified assumption on the spin-alignment timescale (Sec.~\ref{sec:BHtorque}) but might break down on the longer inspiral time. This effect might be especially relevant in the context of differential accretion \citep{2015MNRAS.451.3941G} because it introduces an overall tendency to equalize the BH masses. Our prescription to link the accretion rate of the circumbinary and the circum-BH discs (Sec. \ref{sec:evol_theta_inspiral}) might also not be appropriate in the low-aspect-ratio regime of AGN discs \citep{2015MNRAS.452.3085Y}. Furthermore, \cite{2016MNRAS.460.1243R} reported a prominent pile-up of material at the edge of the disc cavity for $H/R\lesssim 0.1$, with a consequent suppression of the accretion rate. If confirmed, their results imply a spin alignment time that is $\ssim 10 H/R\ssim 100$ times longer, further exacerbating the relevance of the critical obliquity.

{We implemented a quasi-adiabatic approach where the spin-disc evolution is modeled as a sequence of stationary configurations. Although this is well motivated (Sec.~\ref{sec:BHtorque}; \citealt{1976IAUS...73..225R}),  time-dependent solutions might deviate from our profiles  close to the critical region (cf.~\citealt{2015MNRAS.448.1526N}). The formalism developed in this paper does not capture how the system behaves at/beyond criticality. This will need to be tackled by other means.}

Another important limitation of this work lies in the boundary conditions of our BVP (Sec.~\ref{numsetup}). At the outer boundary, we implicitly assume that the angular momentum of the disc is much larger than the spin of the BH at any point during the inspiral. This issue might have important repercussion for systems which are initially counter-aligned \citep{2005MNRAS.363...49K,2008MNRAS.385.1621K}. A more accurate treatment in which the boundary conditions are derived from the circumbinary disc dynamics could %
 provide an escape route to partly avoid the critical obliquity. At the inner boundary, we assume that the disc lies in the equatorial plane of the BH, which might also limit the solution space \citep{1997MNRAS.285..394I,2002MNRAS.337..706L,2015MNRAS.448.1526N}.  {Apsidal precession might also play a role \citep{2016MNRAS.455L..62N,2019MNRAS.487.4965Z}.}

{Finally, we used an effective fluid disc theory \citep{1999MNRAS.304..557O,2013MNRAS.433.2403O} to parametrizes the internal stress driven by the magnetorotational instability. Magnetohydrodynamics simulations in the context of the Bardeen-Petterson effect \citep{2013ApJ...777...21S,  2014ApJ...796..103M,2018ApJ...866....5H,2019ApJ...878..149H,2019arXiv190105970L} reveal a richer phenomenology that is not capture by our approach. Our scheme, however, is computationally cheap and allows for large parameter-space explorations.}

\subsection{Outlook}

Our analysis is an important stepping stone toward predicting the spin angle with which supermassive BHs leave their disc-assisted migration and enter the gravitational-wave driven inspiral. The decoupling of the disc and the binary takes place when the the rate of angular-momentum dissipation through gravitational waves matches the disc viscous timescale. This  corresponds to the orbital separation (\citealt{2014PhRvD..89f4060G})
\begin{align}
R_{\rm dec} &\simeq 3\times 10^{-4} 
\left(\frac{M+M_\star}{2\!\times\!10^7 M_\odot} \right)
\left[\frac{4 M M_\star}{(M+M_\star)^2} \right]^{2/5} 
\notag \\ &\times 
\left(\frac{H/R}{0.002} \right)^{-4/5} 
\left(\frac{\alpha}{0.2} \right)^{-2/5} {\rm pc}\,.
 \end{align}
From this point on, the spins directions change  because of relativistic spin-spin and spin-orbit couplings.
 
Predicting the BH spin orientations at the onset of the gravitational-wave driven regime has important consequences for the  LISA space mission. If spins remain misaligned until merger, the amplitude of the emitted waves will present characteristic precessional modulations~\citep{1994PhRvD..49.6274A}. The inverse problem is even more intriguing: the detection of spin precession with LISA  might provide a leverage to constraint the effectiveness of the Bardeen-Petterson effect and measure the impact of accretion discs on the lives of supermassive BH binaries. Post-merger gravitational recoils also crucially depend on the spin directions, with important repercussions for the occupation fraction of supermassive BHs in their host galaxies~\citep{2007ApJ...667L.133S,2015MNRAS.446...38G}. These lines of investigations will be addressed in future work.

\section*{Acknowledgments}

We thank Gordon Ogilvie for sharing his code to evaluate the viscosities and clarifying many aspects of warp dynamics.  We thank Lorenzo Pino, Giuseppe Lodato, Enrico Ragusa, Rebecca Nealon, Harald Pfeiffer, Chris Nixon, J.J.~Zanazzi, and Julian Krolik for discussions.
D.G. is supported by Leverhulme Trust Grant No. RPG-2019-350. 
G.R. is supported by the VENI  research program with project number 016.Veni.192.233, which is partly financed by the Dutch Research Council (NWO). This work was supported by a STSM Grant from the European COST Action CA16104 ``GWverse''.
R.B. is supported by the University School for Advanced Studies (IUSS) Pavia and the FIP Distinguished Student Award of the American Physical Society.
Computational work was performed on the University of Birmingham BlueBEAR cluster, the Athena cluster at HPC Midlands+ funded by EPSRC Grant No. EP/P020232/1, and the Maryland Advanced Research Computing Center (MARCC).

\bibliographystyle{mnras_tex_edited}
\bibliography{residualalignment}

\end{document}